\documentclass[aps, prl, reprint, floatfix, superscriptaddress, nobibnotes]{revtex4-2}
\usepackage[utf8]{inputenc}
\usepackage{graphicx}
\usepackage{hyperref}
\usepackage{amssymb}
\usepackage{amsmath}
\usepackage{microtype}
\usepackage{color}
\usepackage[caption=false, justification=centerlast]{subfig}
\usepackage[export]{adjustbox}
\usepackage{setspace}
\usepackage{siunitx}

\usepackage{newtxtext}
\usepackage[smallerops]{newtxmath}

\hypersetup{
    colorlinks = true,
    linkcolor  = blue,
    citecolor  = blue,
    urlcolor   = blue,
}
\DeclareMathOperator{\Tr}{Tr}
\renewcommand{\vec}[1]{{\mathbf{#1}}}
\newcommand{\U}[1]{$\text{U}(#1)$}
\newcommand{\Oh}[1]{$\text{O}(#1)$}
\newcommand{\msf}[1]{\mathsf{#1}}

\makeatletter
\g@addto@macro\bfseries{\boldmath}
\makeatother

\begin{document}

\title{Spectroscopic fingerprints of gapless type-II fracton phases}
\author{Oliver Hart}
\email{oliver.hart-1@colorado.edu}
\affiliation{T.C.M. Group, Cavendish Laboratory, JJ Thomson Avenue, Cambridge CB3 0HE, United Kingdom}
\affiliation{Department of Physics and Center for Theory of Quantum Matter, University of Colorado, Boulder, Colorado 80309, USA}
\author{Rahul Nandkishore}
\affiliation{Department of Physics and Center for Theory of Quantum Matter, University of Colorado, Boulder, Colorado 80309, USA}
\date{July 2021}

%%%%%%%%%%%%%%%%%%%%%%%%%%%%%%%%%%%%%%%%%%%%%%%%%%%%%%%%%%%%%%%%%%%%%
%                            ABSTRACT                               %
%%%%%%%%%%%%%%%%%%%%%%%%%%%%%%%%%%%%%%%%%%%%%%%%%%%%%%%%%%%%%%%%%%%%%

\begin{abstract}%
\setstretch{1.1}%
Fracton phases feature elementary excitations with fractionalized mobility and are exciting interest from multiple areas of theoretical physics. However, the most exotic `type-II' fracton phases, like the Haah codes, currently have no known experimental diagnostics. Here, we explain how type-II fracton phases with gapless gauge modes, such as the \U1 Haah code, may be identified experimentally. Our analysis makes use of the `multipole gauge theory' description of type-II fracton phases, which exhibits ultraviolet-infrared (UV-IR) mixing. We show that neutron scattering experiments on gapless type-II fracton phases should generically exhibit exotic pinch points in the structure factor, with distinctive anisotropic contours as a direct consequence of UV-IR mixing. This characteristic pinch point structure provides a clean diagnostic of type-II fracton phases. We also identify distinctive signatures of the $(3+1)$-D \U1 Haah code in the low-temperature specific heat. 
\end{abstract}

\maketitle

%%%%%%%%%%%%%%%%%%%%%%%%%%%%%%%%%%%%%%%%%%%%%%%%%%%%%%%%%%%%%%%%%%%%%
%                          INTRODUCTION                             %
%%%%%%%%%%%%%%%%%%%%%%%%%%%%%%%%%%%%%%%%%%%%%%%%%%%%%%%%%%%%%%%%%%%%%

Quantum spin liquids are exotic phases of matter for which the low-energy effective theory involves fractionalized excitations and deconfined gauge fields (see, e.g., Refs.~\cite{SavaryBalents, ZhouKanodaNg, KnolleFieldGuide} for reviews). Recently, a particularly interesting class of quantum spin liquids known as {\it fracton} phases has been discovered \cite{ChamonCode2005, HaahCode2011}, in which the low-energy elementary excitations exhibit {\it fractionalized mobility} (see Refs.~\cite{NandkishoreHermele, YouPretkoChen} for reviews). In `type-I' fracton phases \cite{VijayHaahFu2}, there exist nontrivial excitations that can move only along {\it subdimensional} manifolds. Such phases are described by symmetric tensor gauge theories \cite{PretkoSub}. Meanwhile, in `type-II' fracton phases \cite{HaahCode2011}, all nontrivial excitations are fully immobile. Type-II fracton phases are described by {\it multipolar} gauge theories \cite{BulmashBarkeshli2018,Schmitz,Gromov2019,Gromov2020duality,WesleiChamon2021}.

Fracton phases are drawing intense interest from multiple areas of theoretical physics, ranging from condensed matter \cite{ChamonCode2005, VijayHaahFu2, PretkoSub, PremPretkoNandkishore, SondhiFracton, HughesFracton, BalentsFracton, PretkoRadzihovsky} to quantum information \cite{BravyiFracton, HaahCode2011, SivaYoshida, recoverable, BernevigFracton, ShirleyChen, ChengFracton} to quantum dynamics \cite{PremHaahNandkishore, PPN, KHN, Sala, IaconisVijayNandkishore, GromovLucasNandkishore, PollmannFracton, MorningstarKhemaniHuse} to high-energy physics \cite{SeibergShao, GorantiaSeiberg}. However,
as far as \emph{experimental} detection of fractons is concerned, 
there is a major outstanding challenge: if a fracton phase were realized in a particular material, how could it be identified in realistic experiments? For gapless $\text{U}(1)$ spin liquids of conventional type, hosting gapless gauge fields and gapped matter, the diagnostic par excellence involves neutron scattering and looking for pinch points \cite{Bramwell2001,Henley2005Powerlaw,Fennell2009,Henley2010annurev}. Similar diagnostics were generalized to type-I fracton phases that host gapless gauge fields in Refs.~\cite{Prem2018PinchPoints, SondhiDiagnostics}, which pointed out that such type-I gapless fracton phases support pinch points with a characteristic symmetry, distinct from the conventional case. Meanwhile, the method of two-dimensional coherent spectroscopy was proposed for the characterization of gapped type-I fracton phases (and also gapped conventional spin liquids) in Ref.~\cite{NandkishoreChoiKim}. However, for the most exotic, type-II fracton phases, we are not aware of a currently known clean experimental diagnostic. 

\begin{figure}[t!]
    \centering
    \includegraphics[width=\linewidth]{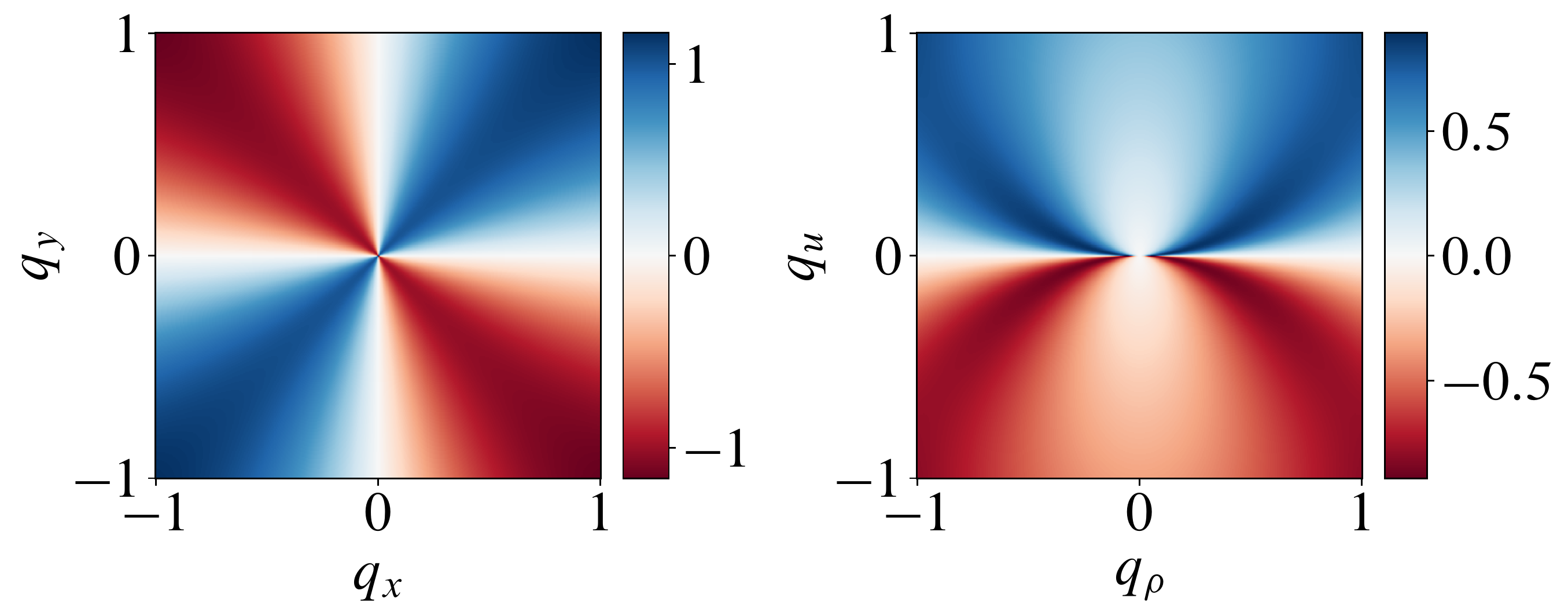}
    \caption{Characteristic pinch point structure of conventional gapless rank-one \U1 spin liquids (left) compared with the pinch points exhibited by gapless type-II theories (right). In the conventional \U1 case, the limit as one approaches $q=0$ depends on the direction through $q_x/q_y$. In contrast, for the \U1 Haah code, the limit depends on $a q_\rho^2/q_u$, where $a$ is a microscopic (lattice) scale, a direct consequence of ultraviolet-infrared (UV-IR) mixing. On the left, we plot $\langle E^x(\vec{q}) E^y(-\vec{q}) \rangle$ in the $q_z=0$ plane at temperature $T=ca^{-1}$, with $c$ the speed of light, and on the right, $\langle E^1(\vec{q}) E^2(-\vec{q}) \rangle$ at the same temperature, where $q_u$ points along [111]. We have plotted here a planar `slice' through three-dimensional momentum space. The slice plotted contains [111], but analogous results are obtained for any slice that is not orthogonal to [111] (see the Supplemental Material~\cite{SM}).}
    \label{fig:pinch-points-schematic}
\end{figure}

In this letter, we identify clean signatures of gapless type-II fracton phases, which should be accessible using existing experimental techniques. Our analysis starts with the `multipolar gauge theory' (MGT) description of type-II fracton phases \cite{BulmashBarkeshli2018,Gromov2019,Gromov2020duality,WesleiChamon2021},
wherein the Gauss law constraint generates fractal charge configurations.
Crucially, all these theories include some degree of ultraviolet-infrared (UV-IR) mixing, with the lattice scale appearing explicitly in the effective field theory description. This is reminiscent
of quantum smectics, where the UV length scale appears naturally as the spacing between layers and which are dual to certain MGTs~\cite{Gromov2020duality}.
We explore the experimental consequences of this UV-IR mixing. We find that the pinch points in neutron scattering once again serve as a concrete experimental diagnostic of the gapless gauge modes.
However, the UV-IR mixing leads to a characteristic pinch point structure in which the contours of the structure factor form
nontrivial power laws, in contrast to the straight lines found in conventional \U1 spin liquids (see Fig.~\ref{fig:pinch-points-schematic}) and symmetric tensor gauge theories.
We believe this unusual structure of pinch points should serve as a generic diagnostic for gapless type-II fracton phases. For the specific case of the $(3+1)$-D \U1 Haah code~\cite{HaahKITP,BulmashBarkeshli2018,Gromov2019,Gromov2020duality}, we also compute the low temperature specific heat capacity, and identify a characteristic $\sim T^2$ scaling behavior.

%%%%%%%%%%%%%%%%%%%%%%%%%%%%%%%%%%%%%%%%%%%%%%%%%%%%%%%%%%%%%%%%%%%%%
%                          FIELD THEORY                             %
%%%%%%%%%%%%%%%%%%%%%%%%%%%%%%%%%%%%%%%%%%%%%%%%%%%%%%%%%%%%%%%%%%%%%

\begin{figure}[t!]
    \centering
    \subfloat[$D_1$]{\includegraphics[width=0.27\linewidth]{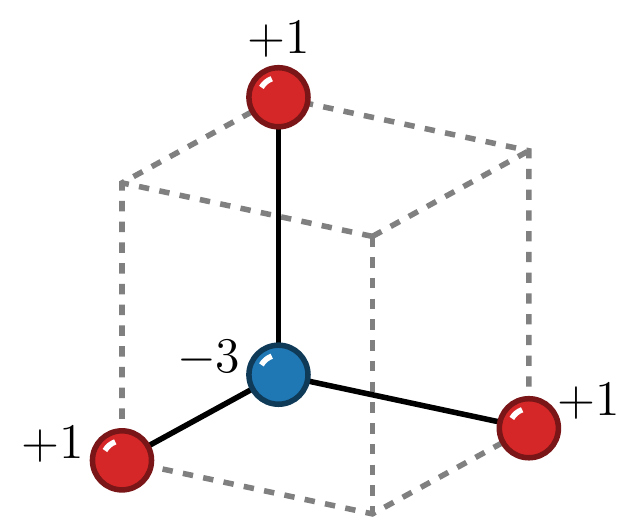}}%
    \hspace{0.015\linewidth}%
    \subfloat[$D_2$]{\includegraphics[width=0.43\linewidth]{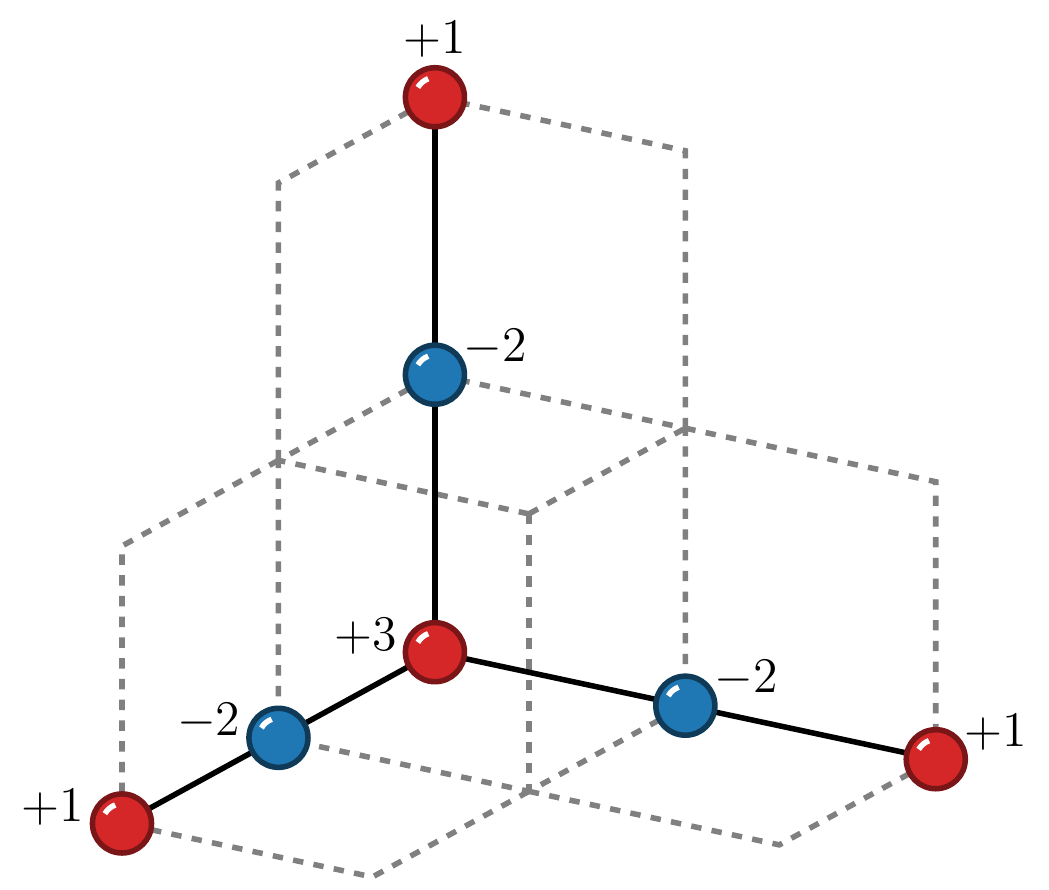}}%
    \hspace{0.015\linewidth}%
    \subfloat[$D_3$]{\includegraphics[width=0.27\linewidth]{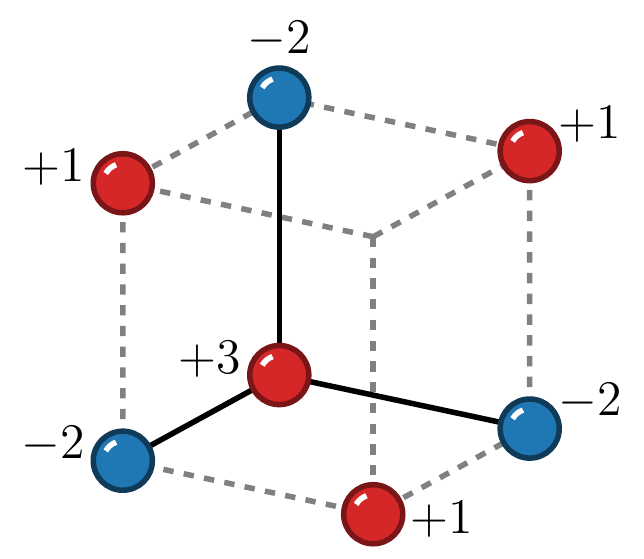}}%
    \caption{Upon discretization, the three differential operators $D_a$ in Eq.~\eqref{eqn:Haah-invariant-derivatives} give rise to three fundamental configurations of charges in the \U1 Haah code. When the \U1 theory is Higgsed, these correspond to the locally creatable charge configurations of the $\mathbb{Z}_2$ Haah code.}
    \label{fig:charge-configurations}
\end{figure}

\textit{Multipole gauge theory}.---%
Here, we review the ingredients necessary to construct MGTs that describe gapless type-II fracton phases such as the \U1 generalizations~\cite{HaahKITP,BulmashBarkeshli2018,Gromov2019,Gromov2020duality} of the Haah code~\cite{HaahCode2011} or the Chamon code~\cite{ChamonCode2005}.
For a more thorough treatment, we refer the reader to the previous literature~\cite{BulmashBarkeshli2018, Gromov2019,Gromov2020duality}. One begins with a set of
generalized derivatives:
\begin{equation}
    D_a = q^i_a \partial_i + q^{ij}_a \partial_i \partial_j + \ldots
    \, ,
\end{equation}
with $a = 1, 2, \ldots$ (unrelated to spatial indices $i,j$).
The dimensionful coefficients $q_a^i, q_a^{ij}, \ldots$ are determined by solving $D_a P_\ell(x)=0$ for a set of polynomials $P_\ell(x)$ with respect to which the system is symmetric.
These constraints on the $D_a$ imply conservation of various linear combinations of multipole moments of the charge density: $\partial_t \int d^d x \, P_\ell(x) \rho(x) = 0$.
This feature represents one of the hallmarks of fractonic phases of matter.
Canonically conjugate electric fields $E_a$ and vector potentials $A_a$ are introduced, which satisfy
\begin{equation}
    E_a = -\dot{A}_a - D_a \Phi
    \, ,
    \label{eqn:electric-fields}
\end{equation}
where $\Phi$ is the scalar potential. The electric fields in Eq.~\eqref{eqn:electric-fields} are invariant under the gauge transformation:
\begin{equation}
    A_a \to A_a + D_a \chi \,, \qquad \Phi \to \Phi - \dot{\chi}
    \, ,
\end{equation}
which is generated by the Gauss law constraint
\begin{equation}
    \sum_a D^\dagger_a E^{\phantom{\dagger}}_a = \rho
    \, ,
    \label{eqn:Gauss-law}
\end{equation}
with $\rho$ a scalar charge density.
In what follows, we will consider the pure gauge theory without charged matter: $\rho=0$.
The differential operators $D^\dagger_a$ appearing in Gauss's law in Eq.~\eqref{eqn:Gauss-law} are related to the $D_a$ via integration by parts $\int d^d x \, f D_a g = \int d^d x \, (D_a^\dagger f) g$~\footnote{The sign has been absorbed into the definition of $D_a^\dagger$, and we have assumed that the fields and their relevant derivatives vanish on the boundaries, if the theory is defined on a manifold with boundaries.}.
Equation~\eqref{eqn:Gauss-law} generalizes the familiar $\partial_i E^i = \rho$ from conventional \U1 electromagnetism (EM).
Importantly, the Gauss law constraint in Eq.~\eqref{eqn:Gauss-law} involves derivatives of \emph{different} degrees when describing type-II fracton theories.
This follows from the fractal configurations of charges that must be created in such theories, as depicted in Fig.~\ref{fig:charge-configurations}.
The system is finally endowed with a Maxwell-like Lagrangian density:
\begin{equation}
    \mathcal{L} = \frac{\epsilon}{2} \sum_a E_a^2 - \frac12 \sum_{a < b} B_{a b}^2 
    \, ,
    \label{eqn:Lagrangian-density-Maxwell}
\end{equation}
where $B_{a b} = D_a A_b - D_b A_a$ are the gauge-invariant magnetic fields.
In general, however, there may be additional gauge invariant quantities (``magnetic fields'') that should be added to the most general quadratic Lagrangian.
We will see that this is indeed the case for the \U1 Haah code, where there exists a nonlinear relationship among the invariant derivatives. Note that we will be working throughout with {\it noncompact} MGTs.

For the (3+1)-D \U1 Haah code~\cite{HaahKITP, BulmashBarkeshli2018,Gromov2019}, there are in general three invariant derivatives~\cite{Gromov2019}
\begin{equation}
    D_1 = \sum_i \partial_i \, , \quad D_2 = \ell_2 \sum_i \partial_i^2 \, , \quad D_3 = \ell_3 \sum_{i < j} \partial_i \partial_j
    \, ,
    \label{eqn:Haah-invariant-derivatives}
\end{equation}
where $i,j,\ldots \in \{x, y, z\}$, and $\ell_2$, $\ell_3$ are UV (lattice) scale quantities with dimensions of length.
To lighten the notation, we will set the dimensionful lengths equal to one another $\ell_2 = \ell_3 \equiv a$, but there is no symmetry that enforces this restriction in general.
There are three magnetic fields of the form $B^a = \epsilon^{a b c}D_b A_c$, two of which are independent since $D_a B^a = 0$.
However, one may also use the nonlinear relationship $a D_1^2 = D_2 + 2D_3$ to construct a further gauge-invariant quantity that is more relevant in the renormalization group sense.
Specifically, we may introduce a fourth ``magnetic field'' $B^4 = D_1 A^1 - a^{-1}(A^2 + 2A^3)$.
This new gauge-invariant quantity is not independent from the other magnetic fields, and there exist various differential relationships that uniquely determine $B^4$ in terms of any two of the $B^a$, as shown in Ref.~\cite{Gromov2019}.

In the absence of matter, the time evolution of $\Phi$ is completely determined by the three fields $A^a$ through Gauss's law in Eq.~\eqref{eqn:Gauss-law}.
We can also choose to work in the analog of the Coulomb (radiation) gauge, in which $D_a^\dagger A^a = 0$, and $\Phi=0$ (see the Supplemental Material (SM)~\cite{SM} for a discussion of different gauge choices).
First, excluding any contribution from $B^4$, the two independent components of the vector potential give rise to two degenerate, gapless degrees of freedom: the `photon,' with two polarizations.
In the Coulomb gauge, the vector potential satisfies
\begin{equation}
    \left( \epsilon \partial_0^2 + D_a^\dagger D^a \right) A^b = 0
    \, ,
\end{equation}
which admits wavelike solutions $A^a(\vec{q}) \exp(i\vec{q}\cdot \vec{x} - i\omega t)$ with the dispersion relation~\cite{BulmashBarkeshli2018}:
\begin{equation}
    \epsilon \omega^2(\vec{q}) = \left(\sum_i q_i\right)^2 + \left( a \sum_{i} q_i^2 \right)^2 + \left( a \sum_{i<j} q_i q_j \right)^2
    \, .
    \label{eqn:dispersion-cartesian}
\end{equation}
The salient features of this dispersion relation become more transparent when it is expressed in terms of a coordinate system aligned with the [111] direction.
Indeed, the Lagrangian defined by the invariant derivatives in Eq.~\eqref{eqn:Haah-invariant-derivatives} is rotationally symmetric about [111].
Let $\vec{q} = q_u \hat{\vec{e}}_u + q_\rho \hat{\vec{e}}_\rho$, where $\hat{\vec{e}}_u$ is aligned with [111], and $\hat{\vec{e}}_\rho$ points radially outwards in the (111) plane.
In this basis, the dispersion relation in Eq.~\eqref{eqn:dispersion-cartesian} reads
\begin{equation}
    \epsilon \omega^2(\vec{q}) = 3 q_u^2 + a^2(q_u^2 + q_\rho^2)^2 + a^2(q_u^2 - \tfrac12 q_\rho^2)^2
    \, ,
    \label{eqn:dispersion-111}
\end{equation}
which is linear, $\omega(\vec{q}) = c|q_u|$, along [111], while it is quadratic, $\omega(\vec{q}) \propto c a q_\rho^2$, in the plane $q_u = 0$.

Upon including a term $\propto -m^2 B_4^2$ in the Lagrangian density in Eq.~\eqref{eqn:Lagrangian-density-Maxwell}, one of the two degenerate photon polarizations develops a gap $\propto m$, while the other is left unaffected and keeps the
dispersion relation in Eq.~\eqref{eqn:dispersion-111} (see the SM for further details~\cite{SM}).
Hence, there is just one nondegenerate, gapless photon branch in general.

%%%%%%%%%%%%%%%%%%%%%%%%%%%%%%%%%%%%%%%%%%%%%%%%%%%%%%%%%%%%%%%%%%%%%
%                          PINCH POINTS                             %
%%%%%%%%%%%%%%%%%%%%%%%%%%%%%%%%%%%%%%%%%%%%%%%%%%%%%%%%%%%%%%%%%%%%%

\textit{Pinch points}.---%
In magnetic systems whose microscopic degrees of freedom are spins, the structure factor $S^{\alpha\beta}(\vec{q}) = \langle S^\alpha (\vec{q}) S^\beta(-\vec{q})\rangle$ can be accessed in
neutron scattering experiments.
While the precise relationship between electric field correlators $\langle E^a(\vec{q}) E^b(-\vec{q}) \rangle$ and the structure factor depends on nonuniversal details, it is expected on
general grounds that the long-distance 
behavior of the spin correlations is determined by the correlation functions of the gapless gauge modes~\cite{Prem2018PinchPoints}.
In this way, the structure factor inherits the putative singular behavior arising in the electric field correlators~\footnote{As shown in the Supplemental
Material~\cite{SM}, the magnetic field correlation functions do not give rise to pinch points. Hence, any projection of 
the microscopic degrees of freedom onto the magnetic fields will not contribute to the pinch point structure.} at long wavelengths.

In conventional \U1 EM (and in spin liquids such as quantum spin ice~\cite{MoessnerSondhiRVB,HermeleQSI2004,BentonSikoraShannon}), the Maxwell Lagrangian density analogous to Eq.~\eqref{eqn:Lagrangian-density-Maxwell} is
\begin{equation}
    \mathcal{L} = \frac12\left( E_i E^i - B_i B^i \right)
    \, ,
\end{equation}
where $B^i = \epsilon^{ijk}\partial_j A_k$. Absent matter, Gauss's law takes the form $\partial_i E^i = 0$, implying that the three spatial components of the electric field are not independent. At strictly zero temperature, the electric field correlation function is
\begin{equation}
    {\langle E^i(\vec{q}) E^j(-\vec{q}) \rangle}_{T=0} \propto \omega(\vec{q}) \left( \delta^{ij} - \frac{q^i q^j}{q^2} \right)
    \, ,
    \label{eqn:standard-pinch-points}
\end{equation}
where $\omega(\vec{q}) = c q$.
The tensor structure derives from Gauss's law: there are two polarizations of the photon, both of which are orthogonal to the momentum $\vec{q}$.
It may readily be observed that the right-hand side annihilates any vector parallel to $\vec{q}$ and therefore acts as a projector into the low-energy~\footnote{In many microscopic Hamiltonians, Gauss' law arises due to a soft energetic constraint. Violations of this constraint are permitted, but they are energetically costly.},
divergence-free sector.
This projector exhibits a pinch point singularity:
its limit as $q \to 0$ depends on the direction that the origin is approached.

Quantum fluctuations are responsible for the factor $\omega(\vec{q})$, which suppresses the pinch points with respect to classical EM
(and, correspondingly, classical spin liquids, such as spin ice~\cite{Isakov2004Dipolar,Henley2005Powerlaw}).
However, at \emph{any} nonzero temperature $T$, the result in Eq.~\eqref{eqn:standard-pinch-points} is modified according to $\langle E^i(\vec{q}) E^j(-\vec{q}) \rangle_T = \langle E^i(\vec{q}) E^j(-\vec{q}) \rangle_{T=0} \coth[\frac12 \beta \omega(\vec{q})]$.
At length scales much greater than the thermal de Broglie wavelength $\lambda_T \sim c/T$, the
factor $\omega(\vec{q})$ that suppresses the pinch points cancels with the low-energy expansion of $\coth[\frac12 \beta \omega(\vec{q})]$~\cite{BentonSikoraShannon}.
Consequently, at sufficiently low energies:
\begin{equation}
    \langle E^i(\vec{q}) E^j(-\vec{q}) \rangle_T \approx T  \left( \delta^{ij} - \frac{q^i q^j}{q^2} \right)
    \, ,
    \label{eqn:nonzero-T-pinch-points}
\end{equation}
in which the pinch point structure is reinstated, with magnitude proportional to temperature.

For the MGT, we define the quantities $Q^a(\vec{q})$ as the eigenvalues of the differential operators $D^a$, i.e., $Q^a(\vec{q}) = e^{-i\vec{q}\cdot\vec{x}} D^a e^{i\vec{q}\cdot\vec{x}}$.
In this language, the dispersion relation in Eq.~\eqref{eqn:dispersion-111} is simply $\epsilon\omega^2(\vec{q}) =  \bar{Q}_a(\vec{q}) Q^a(\vec{q})$.
The analog of the divergence-free condition in the MGT is $D^\dagger_a E^a = 0$.
The Coulomb gauge constraint enforces that the photon polarization ``vector'' $\xi(\vec{q})$ is orthogonal to $Q(\vec{q})$, i.e., $(Q, {\xi})=0$, with respect to the inner product $(v, w) \equiv \bar{v}_a w^a$.
Inclusion of the $mB^2_4$ term in the effective Lagrangian gaps out the photon branch with polarization $\xi_1(\vec{q}) = (a Q^1(\vec{q}), 1, 2)$.
This leaves only one gapless photon with polarization vector $\xi_2(\vec{q})$, defined by orthogonality to both $Q(\vec{q})$ and $\xi_1(\vec{q})$. 
Evaluating the electric field correlator at nonzero temperature $T$, one finds
\begin{equation}
    \langle E^a (\vec{q}) E^b (-\vec{q}) \rangle_T \propto \sum_{r=1}^2  
    \frac{\xi^a_r(\vec{q}) \bar{\xi}^b_r(\vec{q})}{(\xi_r, \xi_r)} \omega_r(\vec{q}) \coth\left[\tfrac12 \beta \omega_r(\vec{q})\right]
    \, ,
    \label{eqn:type-II-correlator}
\end{equation}
where the sum is over the two orthogonal polarizations $r$.
The contribution from the gapped photon branch has no singular behaviour in the vicinity of the origin $q=0$.
Instead, the pinch point structure is determined solely by the gapless photon branch with polarization vector $\xi_2$.
While the indices $a,b,c\ldots$ are not Cartesian, a generic experiment should pick up a contribution from off-diagonal correlators of the type above (as well as contributions from diagonal correlators) and will inherit the singular pinch point structure arising therein. 
The principal features of Eq.~\eqref{eqn:type-II-correlator} are most crisply demonstrated by the off-diagonal correlator $\langle E^1(\vec{q}) E^2(-\vec{q}) \rangle_T$. 
The correlator has the following contribution from the gapless photon branch:
\begin{equation}
    \langle E^1(\vec{q}) E^2(-\vec{q}) \rangle_T
    \propto
    \begin{cases}
        T \dfrac{q_u}{q_\rho^2} &\text{ for } q_u \ll a q_\rho^2 \ll q_\rho \, , \\
        T q_u                  &\text{ for } q_u \gg q_\rho \gg a q_\rho^2 \, , \\
        T \dfrac{q_\rho^2}{q_u} &\text{ for } q_\rho \gg q_u \gg a q_\rho^2 \, ,
    \end{cases}
\end{equation}
for sufficiently low energies $\beta \omega_2(\vec{q}) \ll 1$.
The other correlators exhibit analogous behavior~\cite{SM}.
In contrast to pinch points in conventional, Eq.~\eqref{eqn:standard-pinch-points}, and symmetric tensor spin liquids~\cite{Prem2018PinchPoints}, the limiting behavior of the correlator at long wavelengths depends
not (only) on the direction that the origin is approached (i.e., $q_u/q_\rho$) but on the parameter $q_u/(aq_\rho^2)$.
This gives rise to parabolic contours and thence sharp, distinctive features in the plane $q_u=0$, as shown in Fig.~\ref{fig:pinch-points-schematic}, which we refer to as ``needle-like'' singularities.
The more familiar ``bow-tie'' singularities can be observed by plotting the correlation functions in the $(q_u, a q_\rho^2)$ plane.
This unique long-wavelength behavior is a direct consequence of UV-IR mixing, which is necessary in the field theoretic description of type-II fractonic matter~\cite{Gromov2019}. Other models characterized by dispersion relations involving different powers of momentum in different directions will therefore exhibit parallel behavior, although the parabolic contours may be replaced by
higher-order polynomials.

%%%%%%%%%%%%%%%%%%%%%%%%%%%%%%%%%%%%%%%%%%%%%%%%%%%%%%%%%%%%%%%%%%%%%
%                         HEAT CAPACITY                             %
%%%%%%%%%%%%%%%%%%%%%%%%%%%%%%%%%%%%%%%%%%%%%%%%%%%%%%%%%%%%%%%%%%%%%

\textit{Heat capacity}.---%
In addition to the distinctive consequences of UV-IR mixing in the structure factor, we also consider the low-temperature behavior of the heat capacity.
Since charge excitations are gapped, their density, and hence their contribution to the heat capacity, is suppressed exponentially with temperature below their gap.
This applies also to the gapped photon branch.
Therefore, at temperatures that are low with respect to both gaps,
the dominant contribution to the heat capacity will be the single gapless photon mode.
In lattice systems, where the gauge theory described above is emergent, one must also consider the competing contribution that comes from the vibrational modes of the lattice (phonons).
At long wavelengths, the approximate form of the dispersion relation in Eq.~\eqref{eqn:dispersion-111} is
\begin{equation}
    \omega^2(\vec{q}) \approx c^2q_u^2 + \gamma^2 q_\rho^4
    \, ,
    \label{eqn:low-energy-dispersion}
\end{equation}
where, for Eq.~\eqref{eqn:Lagrangian-density-Maxwell}, $c^2 = 3/\epsilon$ and $\gamma^2 = 5a^2/(4\epsilon)$.
However, this behavior is expected at sufficiently long wavelengths for rather general quadratic Lagrangians (see SM~\cite{SM}).
The heat capacity per unit volume $C_V$ is then given in terms of the dispersion relation as
\begin{equation}
    C_V \propto
    \frac{\partial}{\partial T} \int \frac{{d}^3 q}{(2\pi)^3} \, \frac{\omega(\vec{q})}{\exp\left[\frac{\omega(\vec{q})}{T}\right] - 1}
    \, .
\end{equation}
We may now introduce a new integration variable $\rho=q_\rho^2$ and write the measure as $4\pi dq_u d\rho$.
In terms of the new variables $q_u$ and $\rho$, it is clear that the heat capacity behaves like that of a \emph{two}-dimensional system with a linear dispersion $\omega(\vec{q}) \sim |q|$.
We find that
\begin{equation}
    C_V \propto
    \frac{3\zeta(3)}{4\pi \gamma c} T^2
    \, ,
    \label{eqn:C_v-low-temp}
\end{equation}
with $\zeta(x)$ the Riemann zeta function, interposed between three-dimensional systems with linear or quadratic dispersions, which satisfy $C_V \propto T^3$ and $C_V \propto T^{3/2}$, respectively.
The proportionality in Eq.~\eqref{eqn:C_v-low-temp} is approximate but becomes asymptotically exact for sufficiently low temperatures. Note
that the combination of UV-IR mixing and emergent rotational invariance gives rise to an interesting manifestation of dimensional reduction. Note also that the $\sim T^2$ contribution to heat capacity from the gauge field should dominate over the contribution from acoustic phonons ($\sim T^3$) at low temperatures, making it (in principle) easy to see experimentally. 
While the low temperature $\sim T^2$ behavior hinges on the specific form of the dispersion
relation in Eq.~\eqref{eqn:low-energy-dispersion}, more general anisotropic dispersions will generically produce a heat capacity that
cannot appear in theories where the dispersion relation involves the same (integer) power of
momentum in all directions.

%%%%%%%%%%%%%%%%%%%%%%%%%%%%%%%%%%%%%%%%%%%%%%%%%%%%%%%%%%%%%%%%%%%%%
%                           O(3) ROTORS                             %
%%%%%%%%%%%%%%%%%%%%%%%%%%%%%%%%%%%%%%%%%%%%%%%%%%%%%%%%%%%%%%%%%%%%%

\begin{figure}
    \centering
    \includegraphics[width=0.6\linewidth]{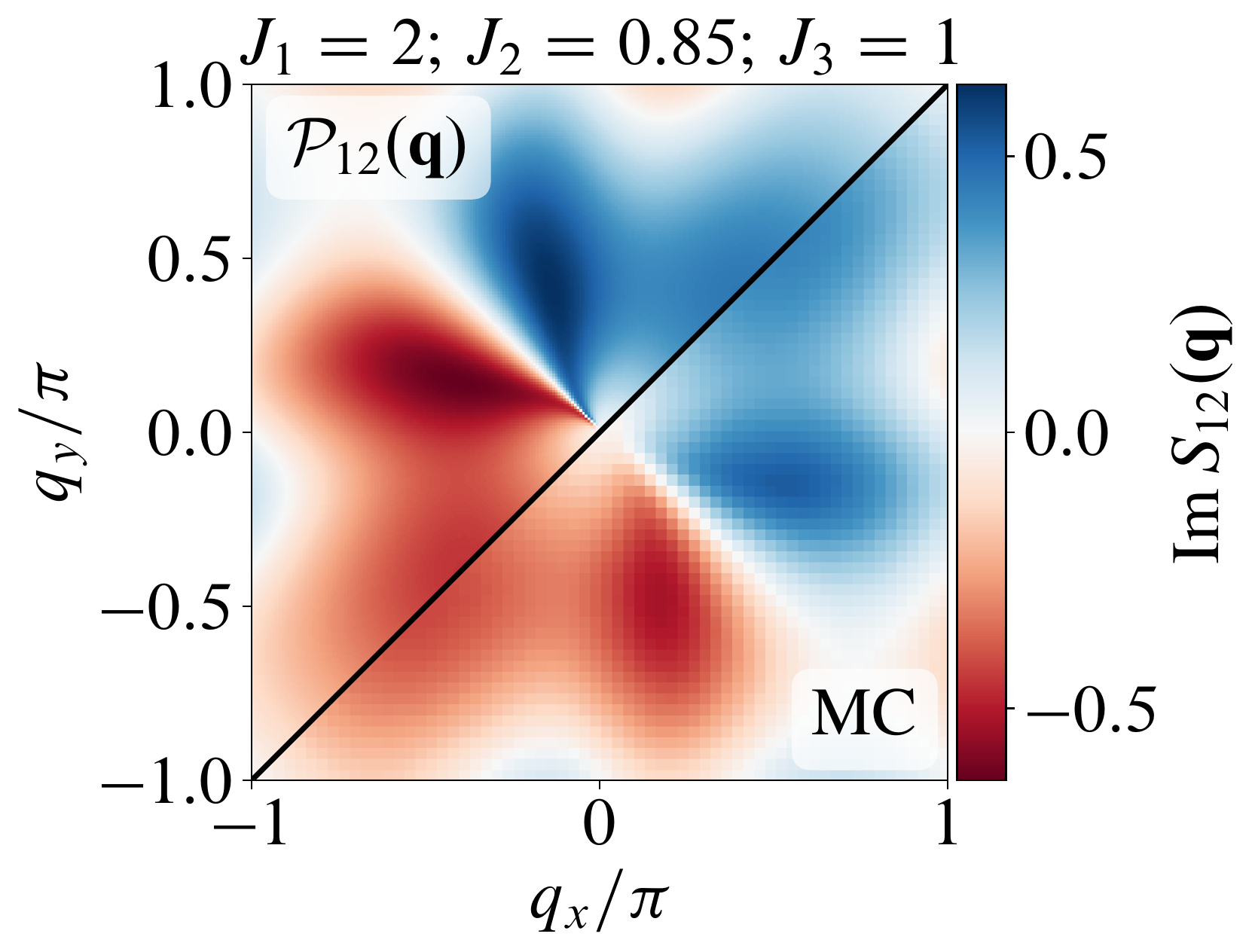}
    \caption{Comparison between the projection formula in Eq.~\eqref{eqn:projection-approx} (upper left) and Monte Carlo data (lower right) for the classical Hamiltonian in Eq.~\eqref{eqn:Hamiltonian-O(3)}. The simulations are carried out for a system of size $\num{98304}$ spins at a temperature $T=0.8$; further details are given in the SM~\cite{SM}.}
    \label{fig:MC-results}
\end{figure}

\textit{Classical spin model}.---%
We can also write down a classical, local, translationally invariant spin
Hamiltonian that exhibits the
same parabolic pinch points as the \U1 Haah code. The model is composed
of \Oh3 rotor variables (unit length Heisenberg spins) $\vec{S}_\vec{r}^{(n)}$.
There are three spins ($n=1,2,3$) per site $\vec{r}$ of the cubic lattice, and
\begin{equation}
    H = \sum_{\vec{r}} \left[ \sum_{n, \vec{r}'} \eta_{\vec{r}\vec{r}'}^{(n)} \vec{S}_{\vec{r}'}^{(n)} \right]^2 \equiv \sum_\vec{r} \vec{M}^\text{tot}_\vec{r} \cdot \vec{M}^\text{tot}_\vec{r}
    \, ,
    \label{eqn:Hamiltonian-O(3)}
\end{equation}
where the coefficients $\eta_{\vec{r}\vec{r}'}^{(n)} = J_n \hat{\eta}^{(n)}_{\vec{r}-\vec{r}'}$, with $\hat{\eta}^{(n)}_{\vec{r}-\vec{r}'}$ chosen to match the
elementary charge configurations depicted in Fig.~\ref{fig:charge-configurations}.
The term inside the square brackets defines $\vec{M}^\text{tot}_\vec{r}$, and
the ground states of the Hamiltonian satisfy the local constraint $\vec{M}^\text{tot}_\vec{r} = \vec{0}$ for all sites $\vec{r}$.
This constraint on ground state spin configurations is the lattice analog of the Gauss law constraint in Eq.~\eqref{eqn:Gauss-law}.
Since the coefficients $\eta_{\vec{r}\vec{r}'}^{(n)}$ satisfy $\sum_{\vec{r}'} \eta_{\vec{r}\vec{r}'}^{(n)} = 0$, the ground state
manifold includes all ferromagnetic spin configurations.
When the dot product in Eq.~\eqref{eqn:Hamiltonian-O(3)} is expanded, the Hamiltonian consists of interactions of the form $\vec{S}_{\vec{r}}^{(m)}\cdot \vec{S}_{\vec{r}'}^{(n)}$, where $\vec{r}$ and $\vec{r}'$ are separated by a maximal (Manhattan) distance of four.
In the language of Ref.~\cite{benton2021topological}, we can write the constraint $\vec{M}_\vec{r}^\text{tot}=\vec{0}$ in Fourier space as
\begin{equation}
    \sum_{n=1}^3 \bar{L}^{(n)}(\vec{q}) \vec{S}^{(n)}(\vec{q}) = \vec{0} \quad \forall \, \vec{q}
    \, ,
    \label{eqn:k-space-Gauss}
\end{equation}
where the $L^{(n)}(\vec{q})$ are given by Fourier transforming the coefficients $\eta^{(n)}_{\vec{r}\vec{r}'}$. The form of the low-temperature spin correlation function is then given approximately by projecting out
the ``vector'' with components $L^{(n)}(\vec{q})$:
\begin{equation}
    S_{mn}(\vec{q}) \equiv \langle \vec{S}^{(m)}(-\vec{q}) \cdot \vec{S}^{(n)}(\vec{q}) \rangle \simeq \frac{1}{\kappa} \mathcal{P}_{mn}(\vec{q})
    \, ,
    \label{eqn:projection-approx}
\end{equation}
where $\kappa$ enforces the sum rule on $S_{mn}(\vec{q})$. The approximation only respects the unit length constraint on average and is strictly valid in the large-$N$ limit, where $N$ is the number of spin components~\cite{Benton2016spin,benton2021topological}.
Monte Carlo simulations of the classical Hamiltonian in Eq.~\eqref{eqn:Hamiltonian-O(3)} are shown in Fig.~\ref{fig:MC-results} and are contrasted with the projection approach in Eq.~\eqref{eqn:projection-approx}.
We observe that the Hamiltonian in Eq.~\eqref{eqn:Hamiltonian-O(3)} indeed exhibits parabolic pinch points at $q=0$ due to the presence of mixed derivatives in the Gauss law constraint in Eq.~\eqref{eqn:k-space-Gauss}.

%%%%%%%%%%%%%%%%%%%%%%%%%%%%%%%%%%%%%%%%%%%%%%%%%%%%%%%%%%%%%%%%%%%%%
%                           DISCUSSION                              %
%%%%%%%%%%%%%%%%%%%%%%%%%%%%%%%%%%%%%%%%%%%%%%%%%%%%%%%%%%%%%%%%%%%%%

\textit{Discussion}.---%
We have identified crisp signatures of gapless type-II fracton phases in pinch points and also in the heat capacity. While our calculations are specifically for the U(1) Haah code in three spatial dimensions, we expect the results to be generic. The key signature of parabolic rather than linear contours around the pinch points is a direct consequence of UV-IR mixing in the `mixed derivative' Gauss law, which is expected to be a generic feature of all type-II theories, although in the generic case, the contours could be arbitrary higher-order polynomials instead of parabolae. We have also shown that analogous features arise in classical spin systems with UV-IR mixing via mixed derivative Gauss law constraints.

We have, however, made three significant assumptions. Firstly, we have treated the MGT as noncompact, while an emergent gauge theory in some, e.g., frustrated magnet is likely to be {\it compact}. Working out the consequences of instantons in a compact MGT remains an important open problem. Secondly, we have chosen to work in the sector with vanishing charge density, corresponding to working about the global ground state of the system. Generalizing the analysis to systems with a nonvanishing charge density would necessitate considering the coupling between the gauge theory and charged matter. Working out the full Maxwell equations and the generalized hydrodynamics and exploring the consequences thereof would be a fruitful project for future work. Finally, we have worked directly with the MGT. It would also be worthwhile to explicitly work out the mapping from a microscopic lattice Hamiltonian to the emergent gauge theory, and to determine to precisely which combination of emergent gauge fields a particular experiment couples. However, since the `microscopic' Hamiltonian leading to the emergent gauge theory is nonuniversal, one would want to have some material in mind that might realize a type-II fracton phase as well as an experimental protocol.
The identification of candidate material realizations (and of additional experimental signatures) remains, of course, an important open problem for future work. 

\emph{Acknowledgements}---%
Work by R.N.~was supported by the U.S.~Department of
Energy, Office of Science, Basic Energy Sciences (BES) under Award No. DE-SC0021346.
O.H.~acknowledges support from an Engineering and Physical
Sciences Research Council (EPSRC) studentship.

%%%%%%%%%%%%%%%%%%%%%%%%%%%%%%%%%%%%%%%%%%%%%%%%%%%%%%%%%%%%%%%%%%%%%
%                          BIBLIOGRAPHY                             %
%%%%%%%%%%%%%%%%%%%%%%%%%%%%%%%%%%%%%%%%%%%%%%%%%%%%%%%%%%%%%%%%%%%%%

\bibliographystyle{aipnum4-1}
\bibliography{library}

\cleardoublepage
\newpage

\onecolumngrid
\begin{center}
\textbf{\large Supplemental Material for ``Spectroscopic fingerprints of gapless type-II fracton phases''}
\vskip 0.4cm
{Oliver{\;\,}Hart\textsuperscript{1,\,2}{\;\,}and{\;\,}Rahul{\;\,}Nandkishore\textsuperscript{2}}
\vskip 0.1cm
\textsuperscript{1}{\fontsize{9.5pt}{11.5pt}\selectfont\emph{T.C.M.~Group, Cavendish Laboratory, JJ Thomson Avenue, Cambridge CB3 0HE, United Kingdom}}\\[-0.75pt]
\textsuperscript{2}{\fontsize{9.5pt}{11.5pt}\selectfont\emph{Department of Physics and Center for Theory of Quantum Matter,\\[-0.75pt] University of Colorado, Boulder, Colorado 80309, USA}}\\[-0.75pt]
{\fontsize{9pt}{11pt}\selectfont(Dated: July 2021)}
\vskip 0.75cm
\end{center}
\twocolumngrid

% Prefix 'S' and reset counters
\setcounter{equation}{0}
\setcounter{figure}{0}
\setcounter{table}{0}
\setcounter{page}{1}
\makeatletter
\renewcommand{\theequation}{S\arabic{equation}}
\renewcommand{\thefigure}{S\arabic{figure}}

%%%%%%%%%%%%%%%%%%%%%%%%%%%%%%%%%%%%%%%%%%%%%%%%%%%%%%%%%%%%%%%%%%%%%
%                         SUPPLEMENTARY                             %
%%%%%%%%%%%%%%%%%%%%%%%%%%%%%%%%%%%%%%%%%%%%%%%%%%%%%%%%%%%%%%%%%%%%%

\section{Equations of motion}

In the presence of a nonzero $B_4$ term, the Lagrangian density of the pure multipole gauge theory is
\begin{multline}
    \mathcal{L}( \Phi, \{ A^a \}) = \frac{\epsilon}{2} (-\dot{A}^a - D^a \Phi)^2 - \\
    \frac12 (\epsilon_{a b c} D^b A^c) (\epsilon^{a d e}D_d A_e) - \frac{m^2}{2} B_4^2
    \, .
    \label{eqn:general-Lagrangian-density}
\end{multline}
The ``magnetic fields'' $B^a$ with $a=1,2,3$ are given by $B^a = \epsilon^{a b c} D_b A_c$.
However, there exists a fourth gauge invariant quantity, $B^4 = D_1 A^1 - a^{-1}(A^2 + 2A^3)$, whose gauge invariance follows from the nonlinear relationship amongst the invariant derivatives $a D_1^2= D_2 + 2D_3$.
The equation of motion for the scalar potential $\Phi$, which follows from~\eqref{eqn:general-Lagrangian-density}, gives Gauss' law
\begin{equation}
    D^\dagger_a E^a = 0
    \, .
\end{equation}
Equivalently, one may write Gauss' law in terms of the scalar and vector potentials: $D_a^\dagger D^a \Phi = - \partial_0 D_a^\dagger A^a$.
Therefore, the scalar potential is uniquely determined (up to boundary conditions) by the three components of the vector potential.
In the Coulomb (radiation) gauge, $D^\dagger_a A^a = 0$, we can choose $\Phi=0$ also (at least in the absence of charged matter).
Variation of the action with respect to the three components of the vector potential gives rise to
the generalised Maxwell equations
\begin{align}
    \epsilon \partial_0 E^1 &= -\epsilon^{1 b c} D^\dagger_b B_c + m^2 D_1^\dagger B^4 \\
    \epsilon \partial_0 E^2 &= -\epsilon^{2bc} D^\dagger_b B_c - m^2 a^{-1} B^4 \\
    \epsilon \partial_0 E^3 &= -\epsilon^{3bc} D^\dagger_b B_c - 2 m^2 a^{-1} B^4 \, .
\end{align}
Making use of the Coulomb gauge condition, we can write down the corresponding equations of motion for the components of the vector potential
\begin{align}
    \epsilon \partial_0^2 A^1 &= -D^\dagger_a D^a A^1 - m^2 D_1^\dagger[D_1 A^1 - a^{-1}(A^2 + 2A^3)] \\
    \epsilon \partial_0^2 A^2 &= -D^\dagger_a D^a A^2 +  m^2 a^{-2} [D_1 A^1 - (A^2 + 2A^3)]  \\
    \epsilon \partial_0^2 A^3 &= -D^\dagger_a D^a A^3 + 2m^2 a^{-2} [D_1 A^1 - (A^2 + 2A^3)] \, .
\end{align}
Taking the Fourier transform,
the normal modes and their corresponding frequencies are determined by the eigenvalues and eigenvectors of the Hermitian matrix
\begin{equation}
    \begin{pmatrix}
        |Q|^2 + m^2 |Q^1|^2 & -m^2a^{-1}\bar{Q}^1 & -2m^2 a^{-1} \bar{Q}^1 \\
        -m^2 a^{-1} Q^1 & |Q|^2 + m^2a^{-2} & 2m^2 a^{-2} \\
        -2m^2 a^{-1} Q^1 & 2m^2 a^{-2}  & |Q|^2 + 4m^2 a^{-2}
    \end{pmatrix}
    \, ,
    \label{eqn:normal-mode-matrix}
\end{equation}
where we remind the reader that $Q^a(\vec{q})$ are defined as the eigenvalues of the differential operators $D^a$, i.e., $Q^a(\vec{q}) = e^{-i\vec{q}\cdot\vec{x}} D^a e^{i\vec{q}\cdot\vec{x}}$. The overline denotes complex conjugation.
The matrix~\eqref{eqn:normal-mode-matrix} has one nondegenerate eigenvector 
\begin{equation}
    \xi_1(\vec{q}) = (a Q^1(\vec{q}), 1, 2)^T
    \label{eqn:gapped-eigenvector}
\end{equation}
corresponding to the eigenvalue
\begin{equation}
    \epsilon \omega_1^2(q) = 5m^2a^{-2} + (1+m^2)|Q^1|^2 + |Q^2|^2 + |Q^3|^2
    \, ,
    \label{eqn:normal-mode-freq-gapped}
\end{equation}
which is gapped for nonzero $m$.
The nonlinear relationship amongst the invariant derivatives ensures that $\xi_1(\vec{q})$ satisfies the Coulomb gauge condition
\begin{equation}
    (\xi_1, Q) = a|Q^1|^2 + Q^2 + 2Q^3 = 0
    \, .
\end{equation}
The matrix~\eqref{eqn:normal-mode-matrix} also has a degenerate subspace, orthogonal to $\xi_1(\vec{q})$.
The eigenvector $Q(\vec{q})$ is removed by the Coulomb gauge condition, leaving the vector $\xi_2(\vec{q})$, defined by orthogonality to both $\xi_1(\vec{q})$ and $Q(\vec{q})$.
Explicitly, the orthogonal polarisation vector is
\begin{equation}
    \xi_2(\vec{q}) \propto (2Q^2 - Q^3, 2Q^1-a Q^1Q^3, a Q^1 Q^2 - Q^1)^T
    \, .
    \label{eqn:gapless-eigenvector}
\end{equation}
The frequency of this normal mode is unaffected by the addition of $B_4$ and consequently remains gapless:
\begin{equation}
    \epsilon \omega_2^2(\vec{q}) = \bar{Q}_a Q^a \equiv |Q(\vec{q})|^2
    \, .
    \label{eqn:normal-mode-freq-gapless}
\end{equation}

The Lagrangian density~\eqref{eqn:general-Lagrangian-density} was chosen for its simplicity and to make the analogies with conventional \U1 electromagnetism more transparent. However, it is far from the most general quadratic Lagrangian that can be written down.
To see which features remain qualitatively unchanged in the presence of a more general (quadratic) Lagrangian,
we broaden the above analysis to a generic quadratic form for the magnetic field contribution.
Specifically, the magnetic field contribution is generalised to
$\tfrac12 g_{\mu \nu} B^\mu B^\nu$, where $\mu, \nu = 1, 2, 3, 4$.
The matrix $g_{\mu \nu}$ must be positive definite to ensure that energies are bounded from below with no flat directions (and may be chosen without loss of generality to be symmetric).
The generalisation of~\eqref{eqn:general-Lagrangian-density} is then
\begin{equation}
    \mathcal{L}(\Phi, \{ A^a \}) = \frac{1}{2} \sum_{a=1}^3 \epsilon E_a E^a -
    \frac12 \sum_{\mu, \nu = 1}^4 g_{\mu\nu} B^\mu B^\nu
    \, .
    \label{eqn:mostgeneral-Lagrangian-density}
\end{equation}
\begin{figure*}
    \centering
    \includegraphics[width=0.5\linewidth]{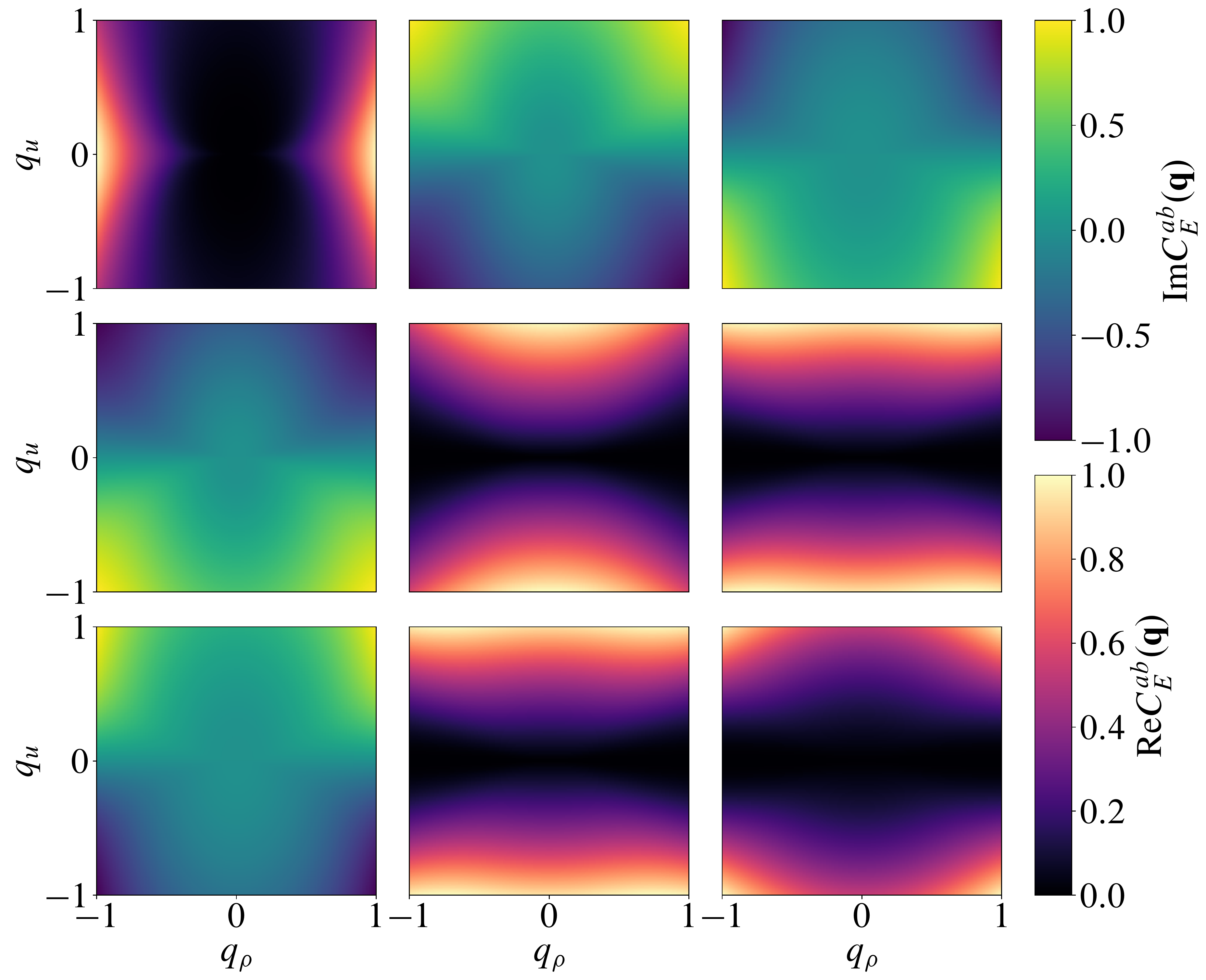}%
    \hfill%
    \includegraphics[width=0.5\linewidth]{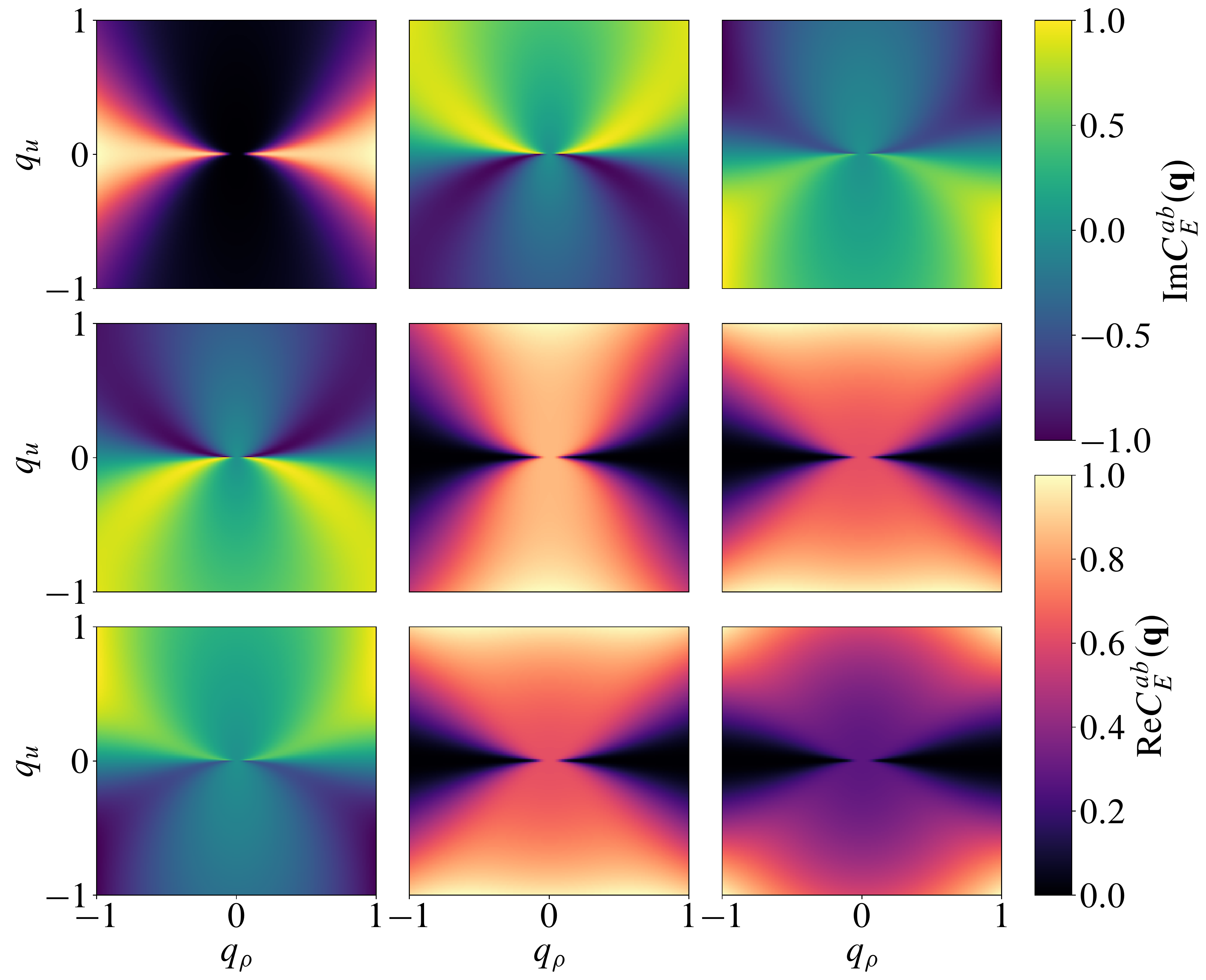}
    \caption{Left: electric field correlation functions at strictly zero temperature, $C_E^{ab}(\vec{q})$, where $a$ and $b$ correspond to the row and column indices, respectively. The factor $\omega(\vec{q})$ suppresses the singular behaviour of the correlation functions near $q=0$. Right: at nonzero temperature, here $T=ca^{-1}$, the needle-like pinch points at long wavelengths can now be observed. In all panels, we plot the contribution from the gapless photon branch only, and normalise by the maximum absolute value of the correlator over the plotted momenta.}
    \label{fig:E-correlation}
\end{figure*}
\begin{figure*}
    \centering
    \includegraphics[width=0.5\linewidth]{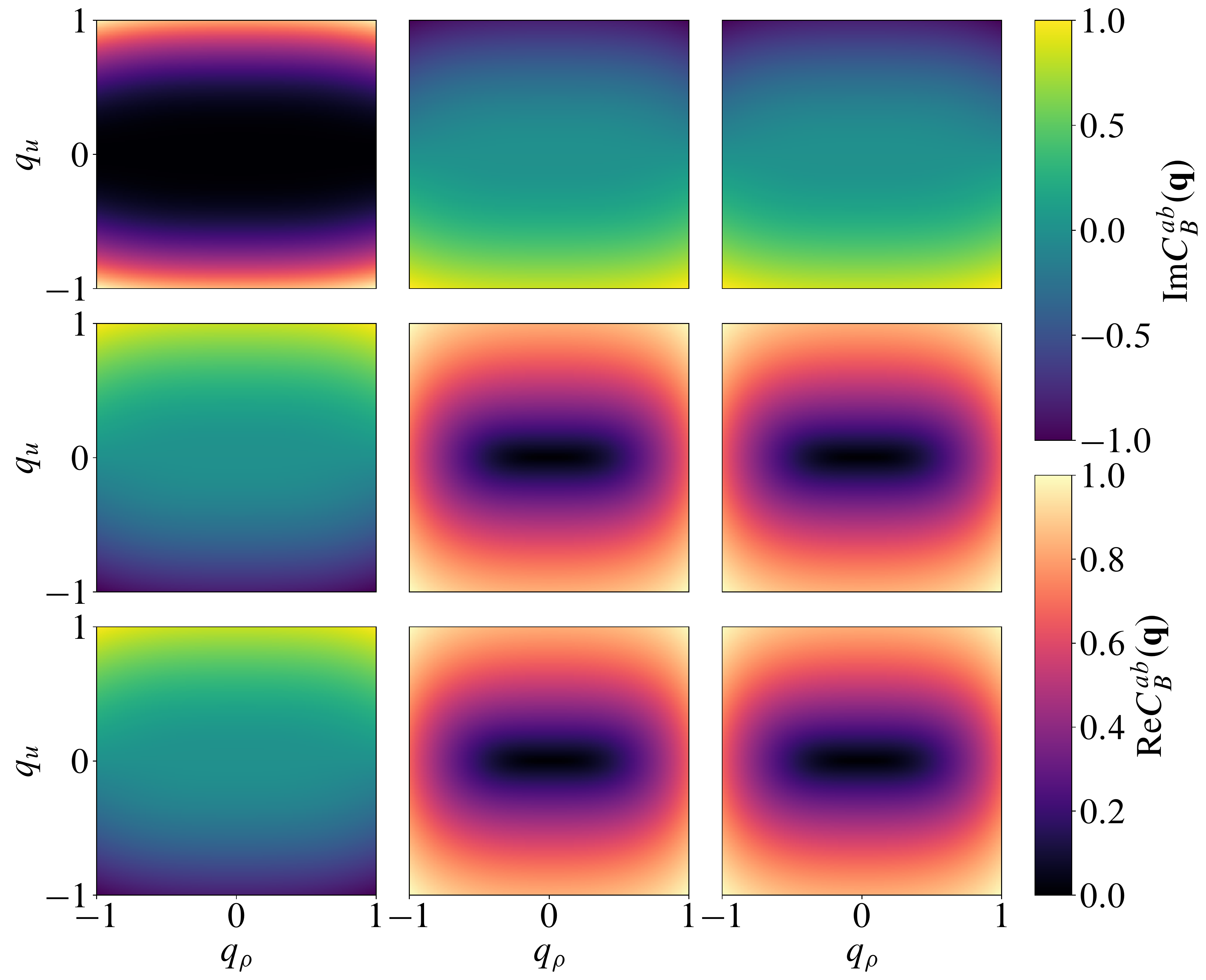}%
    \hfill%
    \includegraphics[width=0.5\linewidth]{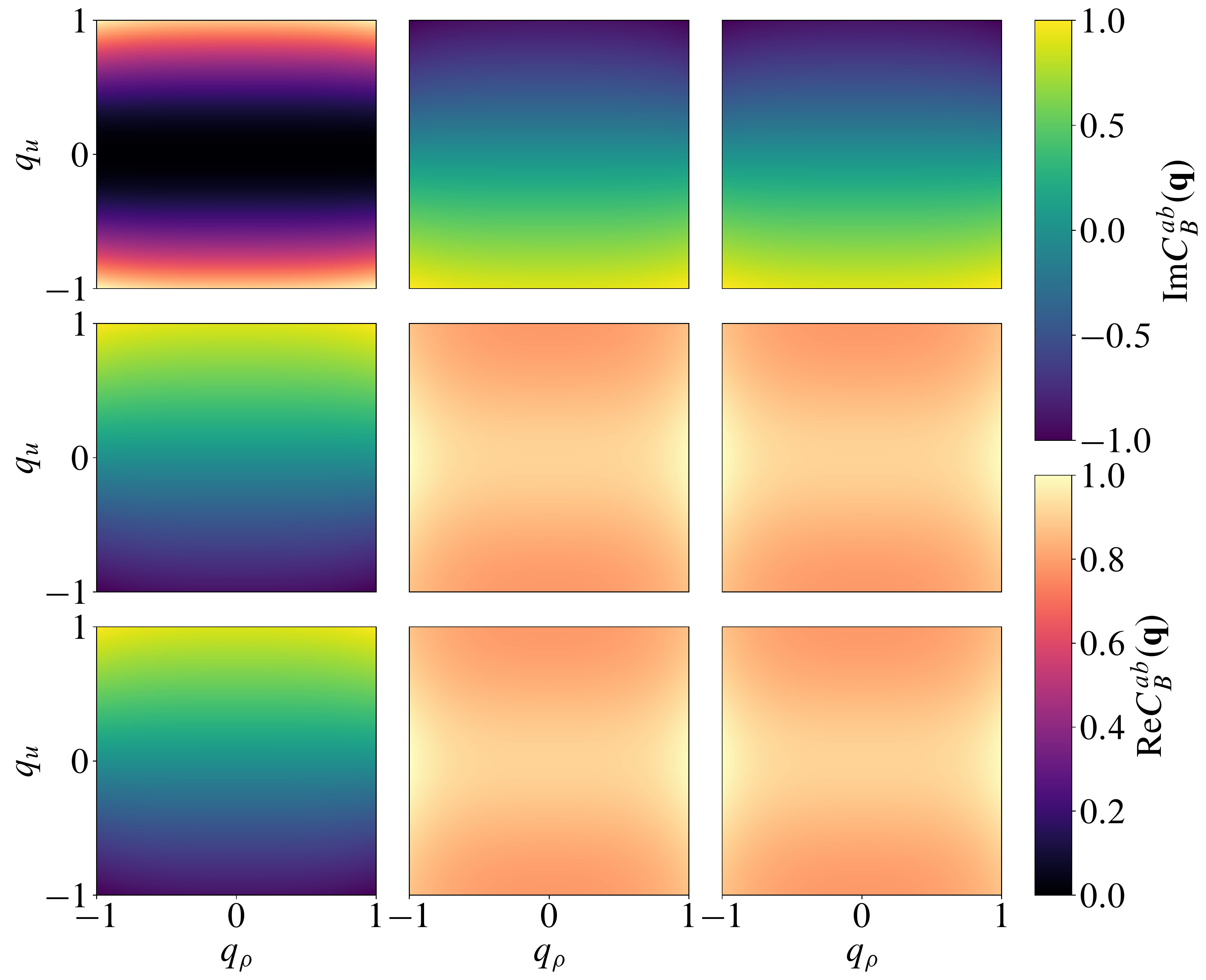}
    \caption{Left: magnetic field correlations at strictly zero temperature, $C_B^{ab}(\vec{q})$, where $a$ and $b$ correspond to the row and column indices, respectively. Right: at nonzero temperature, here $T=ca^{-1}$, there is still no singular behaviour of the correlation functions in the vicinity of the origin. In all panels, we plot the contribution from the gapless photon branch only, and normalise by the maximum absolute value of the correlator over the plotted momenta.}
    \label{fig:B-correlation}
\end{figure*}
\clearpage
In the following, the summation symbols will be omitted. Implicit summation over repeated Greek indices $\mu$, $\nu$ will consistently be over the set $\{1, 2, 3, 4\}$, while summation over repeated indices belonging to the Latin alphabet $a, b$ will be over $\{ 1, 2, 3 \}$.
The magnetic fields are related to the components of the vector potential through a matrix of differential operators
$B^\mu = M[\{ D_b \}]^{\mu}{}_{a} A^a$, where
\begin{equation}
    M[D] =
    \begin{pmatrix}
        0 & -D_3 & D_2 \\
        D_3 & 0 & -D_1 \\
        -D_2 & D_1 & 0 \\
        D_1 & - a^{-1} & -2 a^{-1}
    \end{pmatrix}
    \, .
    \label{eqn:M-matrix}
\end{equation}
In terms of the matrix operator $M$, the equations of motion for $A^a$ can be written compactly as
\begin{equation}
    - \epsilon \partial_0^2 A_a =  M[D]^\dagger_{a}{}^{\mu} g_{\mu\nu} M[D]^{\nu}{}_{b} A^b
    \, ,
\end{equation}
where Hermitian conjugation acts on both matrix and spatial indices (i.e., $(M[D]^\dagger)_{a}{}^{\mu} = M[D^\dagger]^{\mu}{}_{a}$).
In this notation, the transition to Fourier space is simple
\begin{equation}
    \epsilon \omega^2 A_a(\vec{q}) =  M[Q(\vec{q})]^\dagger_{a}{}^{\mu} g_{\mu\nu} M[Q(\vec{q})]^{\nu}{}_{b} A^b(\vec{q})
    \, .
    \label{eqn:normal-modes-evals}
\end{equation}
which reduces the calculation of the normal modes to a standard Hermitian eigenvalue problem.
To show that one photon branch always remains gapless, we perform a singular value decomposition of the matrix $M[Q] = U \Sigma V^\dagger$.
We find that the matrix has the singular values
\begin{align}
    \Sigma_1(\vec{q}) &= \sqrt{5a^{-2} + 2|Q^1|^2 + |Q^2|^2 + |Q^3|^2} \, , \label{eqn:gapped-singular-value} \\
    \Sigma_2(\vec{q}) &= |Q(\vec{q})| \, , \label{eqn:gapless-singular-value}
     \\
    \Sigma_3(\vec{q}) &= 0 \, , \label{eqn:vanishing-singular-value}
\end{align}
equivalent to~\eqref{eqn:normal-mode-freq-gapped} and~\eqref{eqn:normal-mode-freq-gapless} for $m=1$.
The vanishing singular value $\Sigma_3$ corresponds to the right singular vector $Q(\vec{q})=(Q^1, Q^2, Q^3)^T$.
The two-dimensional subspace perpendicular to this vector satisfies the Coulomb gauge condition.
As $q \to 0$, the matrix $M$ has a \emph{second} vanishing singular value
corresponding to Eq.~\eqref{eqn:gapless-singular-value}.
The vectors that make up the columns of $V$ are just the vectors $\xi_r(\vec{q})$ defined previously in Eqs.~\eqref{eqn:gapped-eigenvector} and~\eqref{eqn:gapless-eigenvector}.
We will also introduce the matrix $\tilde{g} = U^\dagger g U$.
Expressing the eigenvalue problem in the basis of right singular vectors,
we arrive at the matrix
\begin{equation}
    \begin{pmatrix}
    \tilde{g}_{11}  \Sigma_1 \Sigma_1  & \tilde{g}_{12} \Sigma_1 \Sigma_2 \\
    \tilde{g}_{21} \Sigma_2 \Sigma_1 & \tilde{g}_{22} \Sigma_2 \Sigma_2 
    \end{pmatrix}
    \, .
    \label{eqn:g-perp}
\end{equation}
The eigenvalues and eigenvectors of this matrix can now be computed with ease.
Near $q=0$, the two eigenvalues of~\eqref{eqn:g-perp} are approximately
\begin{align}
    \Omega_1^2(\vec{q}) &\simeq \tilde{g}_{11} \Sigma_1^2(\vec{q}) \, , \label{eqn:gapped-freq} \\
    \Omega_2^2(\vec{q}) &\simeq \frac{\Sigma_2^2(\vec{q})}{\tilde{g}_{11}} \det \tilde{g}_\perp \, , \label{eqn:gapless-freq}
\end{align}
if $\Sigma_2 \tilde{g}_{22} \ll \Sigma_1 \tilde{g}_{11}$ and $\Sigma_2^2 \tilde{g}_{12}\tilde{g}_{21} \ll \Sigma_1^2 \tilde{g}_{11}^2$,
where $\tilde{g}_\perp = \left(\begin{smallmatrix}\tilde{g}_{11} & \tilde{g}_{12} \\ \tilde{g}_{21}  & \tilde{g}_{22}\end{smallmatrix}\right)$ is the matrix $\tilde{g}_{\mu\nu}$ projected into the subspace orthogonal to $Q$.
Note that since $g_{\mu\nu}$ is positive definite, so is $\tilde{g}$, and hence its diagonal elements are positive, $0<\tilde{g}_{\mu\mu} <\Tr g$ (no summation), and its principal minors are also positive ($\det \tilde{g}_\perp > 0$).
The left singular vectors have a well-defined limit as $q\to 0$, allowing us to write
\begin{gather}
    \tilde{g}_{11}(\vec{0}) = g_{44}  , \quad \tilde{g}_{12}(\vec{0}) = -\frac{1}{\sqrt{5}} \left( g_{42}+2g_{43} \right) \\
    \tilde{g}_{21}(\vec{0}) = -\frac{1}{\sqrt{5}} \left( g_{24}+2g_{34} \right), \quad   \tilde{g}_{22}(\vec{0}) = \frac{1}{5} \left( g_{22} + 4 g_{33} \right)
    \, .
\end{gather}
Therefore $\Omega_2(\vec{q}) \sim \Sigma_2(\vec{q})$ in the long wavelength limit and remains gapless.

If the electric field contribution in Eq.~\eqref{eqn:mostgeneral-Lagrangian-density} is generalised to $\tfrac12 \epsilon_{ab}E^{a}E^{b}$, then Gauss' law is modified to $\epsilon_{ab} {D^a}^\dagger E^b = 0$, and the Coulomb gauge condition is similarly generalised to $\epsilon_{ab} {D^a}^\dagger A^b = 0$. In Fourier space, the equation determining the normal modes is now a generalised eigenvalue problem
\begin{equation}
     \omega^2 \epsilon_{ab} A^b(\vec{q}) =  M[Q(\vec{q})]^\dagger_{a}{}^{\mu} g_{\mu\nu} M[Q(\vec{q})]^{\nu}{}_{b} A^b(\vec{q})
    \, .
\end{equation}
The vector $Q = (Q^1, Q^2, Q^3)$ remains a solution with $\omega=0$. The other two solutions are orthogonal to $Q$ with respect to the metric $\epsilon_{ab}$, and therefore satisfy the generalised Coulomb gauge condition.
Using arguments analogous to those used to derive Eqs.~\eqref{eqn:gapped-freq} and~\eqref{eqn:gapless-freq},
one may show that the frequency of the gapless mode becomes
\begin{equation}
    \Omega_2^2(\vec{q}) \simeq \frac{\Sigma_2^2(\vec{q})}{\tilde{g}_{11} (\epsilon^{-1}_\perp)_{11}} \det \tilde{g}_\perp \det \epsilon^{-1}_\perp
    \, .
\end{equation}
The approximate equality holds for sufficiently long wavelengths, and we have defined $(\epsilon_\perp^{-1})_{rs} = (\xi_r, \epsilon^{-1} \xi_s)$.
Hereafter, we will take the `mass matrix' $\epsilon_{ab}$ to be proportional to the identity, $\epsilon_{ab} = \epsilon \delta_{ab}$.

\begin{figure*}[t]
    \centering
    \includegraphics[height=0.19\linewidth,valign=c]{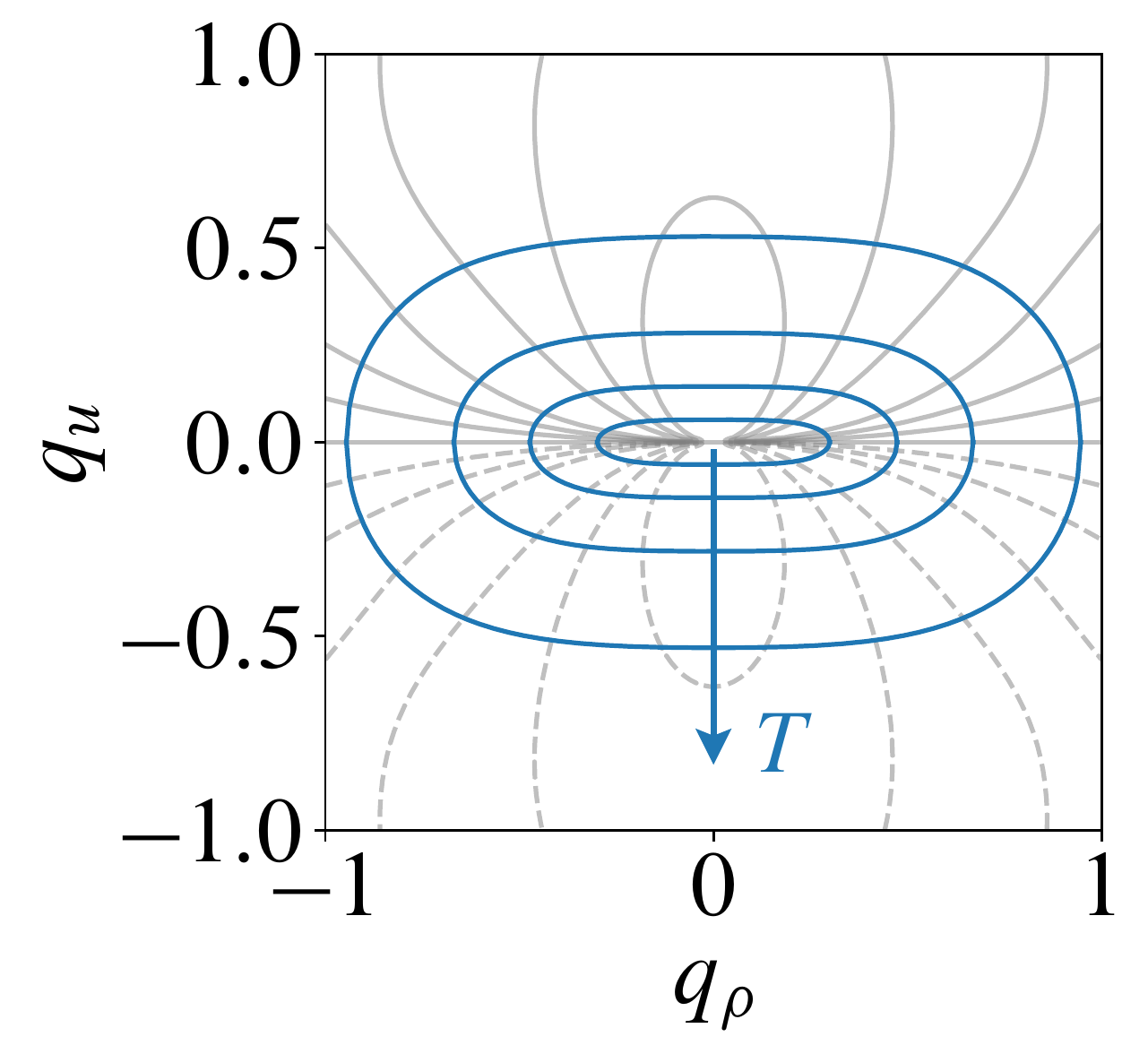}
    \includegraphics[height=0.2\linewidth,valign=c]{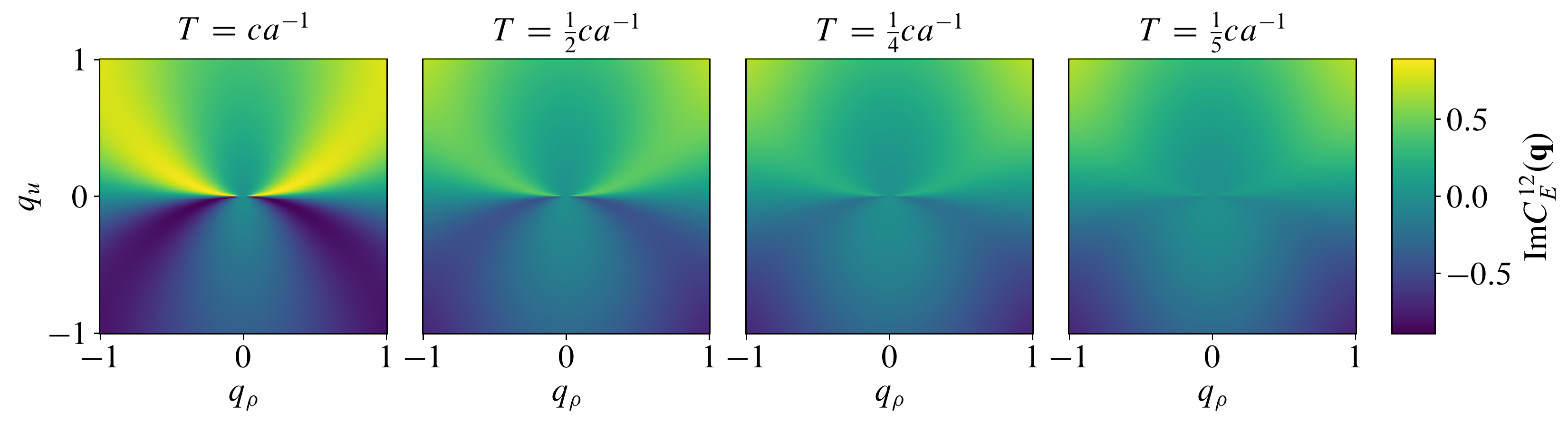}
    \caption{Evolution of the pinch points with temperature. Left: a nonzero temperature $T$ defines the surface $\beta \omega(\vec{q})=1$ in momentum space. Asymptotically, as may be seen from Eq.~(15) in the main text, this surface corresponds to an ellipsoid with a semi-major (minor) axis of length $q_\rho \sim \sqrt{T}$ ($q_u \sim T$). The blue lines correspond to cuts through these surfaces for temperatures $Ta/c=0.1, 0.25, 0.5, 1$, while the gray lines are the contours of $C_E^{ab}(\vec{q})$. Right: the corresponding pinch points. The magnitude of the singular component is $\propto T$, and so the pinch points gradually disappear as temperature is lowered to zero. All plots share the same colour bar.}
    \label{fig:temperature-variation}
\end{figure*}

%%%%%%%%%%%%%%%%%%%%%%%%%%%%%%%%%%%%%%%%%%%%%%%%%%%%%%%%%%%%%%%%%%%%%
%                          QUANTISATION                             %
%%%%%%%%%%%%%%%%%%%%%%%%%%%%%%%%%%%%%%%%%%%%%%%%%%%%%%%%%%%%%%%%%%%%%

\section{Canonical quantisation in the Coulomb gauge}

Since the classical Hamiltonian in the Coulomb gauge corresponds to a collection of simple harmonic oscillators,
the theory can be quantised by introducing raising and lowering operators through the mode expansion
\begin{align}
    A^a(\vec{r}) &= \sum_{\vec{q}} \sum_{r=1}^2   \frac{1}{\sqrt{2V\omega_r(\vec{q})}} \left[\epsilon_r^{a}(\vec{q}) a_{\vec{q}}^r e^{i \vec{q} \cdot \vec{r}} + \text{H.c.} \right] \, , \\
    E^a(\vec{r}) &= -i\sum_{\vec{q}} \sum_{r=1}^2 \sqrt{\frac{\omega_r(\vec{q})}{2V}} \left[\epsilon_r^{a}(\vec{q}) a_{\vec{q}}^r e^{i \vec{q} \cdot \vec{r}} - \text{H.c.} \right] \, ,
\end{align}
where $r=1,2$ are the two photon polarisations, $V$ is the system's volume, and $a_\vec{k}^r$ satisfy the canonical commutation relations, $[a_{\vec{p}}^r , {a_\vec{q}^s}^\dagger] = \delta_{\vec{p} \vec{q}} \delta^{rs}$, if the polarisation vectors $\epsilon_r$ are orthonormal $(\epsilon_r, \epsilon_s) = \delta_{rs}$.
Gauss' law $D^\dagger_a E^a = 0$ and the Coulomb gauge condition $D^\dagger_a A^a = 0$ are satisfied if the photon polarisation vectors are orthogonal to $Q$, i.e., $(Q, \epsilon_r)=0$.
The Hamiltonian is diagonalised if we take the polarisation vectors to be
the normal modes identified in the previous section, $\xi_r$.
Specifically, the Hamiltonian assumes the simple harmonic oscillator form
\begin{equation}
    H = \frac12 \sum_\vec{q}  \sum_{r=1}^{2} \omega_r(\vec{q}) \left( {a_\vec{q}^r}^\dagger a_\vec{q}^r + a_\vec{q}^r {a_\vec{q}^r}^\dagger \right)
    \, ,
    \label{eqn:Coulomb-gauge-Hamiltonian}
\end{equation}
where the photon dispersion relations for the two branches are given by Eqs.~\eqref{eqn:normal-mode-freq-gapped} and~\eqref{eqn:normal-mode-freq-gapless}
[or, more generally, by the eigenvalues of Eq.~\eqref{eqn:normal-modes-evals}].

%--------------------------------------------------------------------
%--------------------------------------------------------------------
%--------------------------------------------------------------------

\begin{figure*}[t]
    \centering
    \includegraphics[height=0.16\linewidth,valign=t]{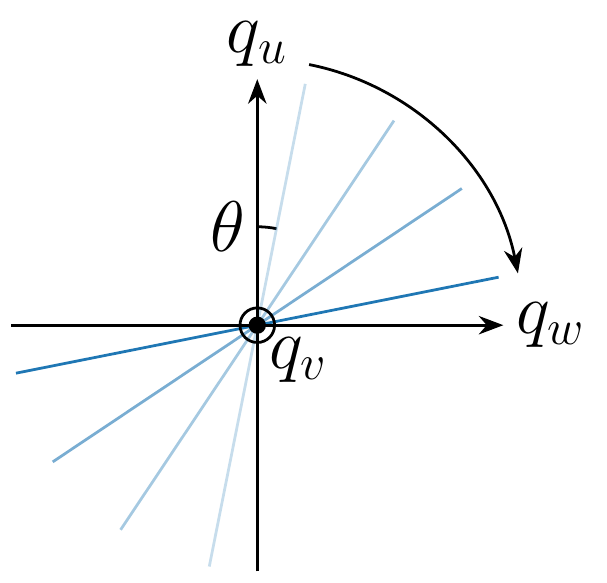}
    \includegraphics[height=0.2\linewidth,valign=t]{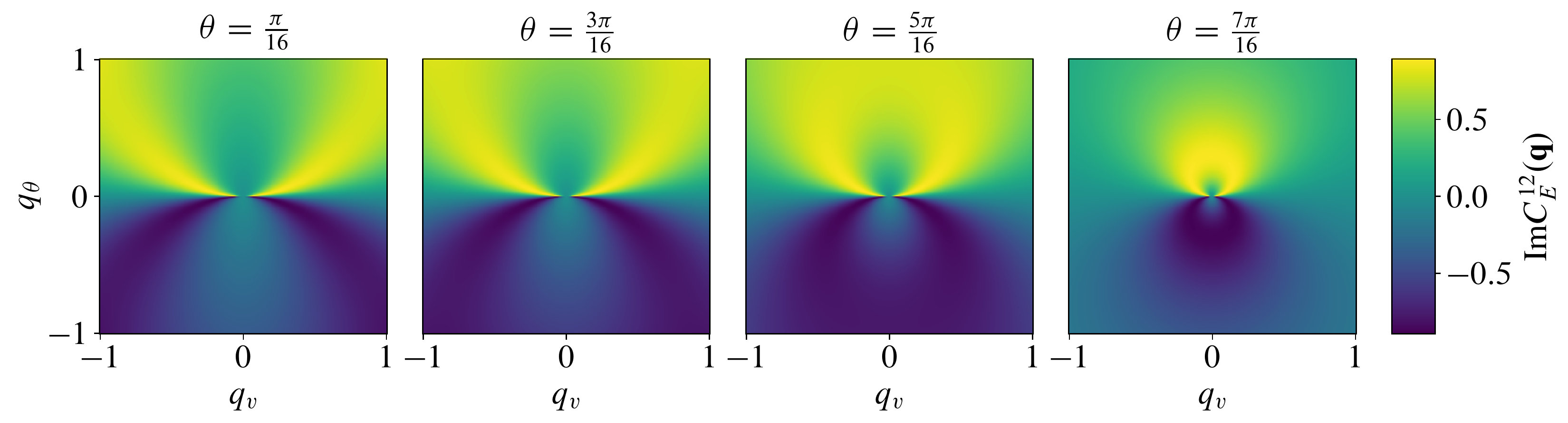}
    \caption{Evolution of the pinch points as the cut through momentum space is rotated by an angle $\theta$ with respect to the [111] direction. Specifically, we plot the correlation function $C_E^{12}(\vec{q})$ in the $q_\theta$-$q_v$ plane, where $q_v$ and $q_w$ are coordinates in the plane orthogonal to [111], and $q_\theta = q_u \cos\theta + q_w \sin\theta$. Parabolic pinch points are always present at sufficiently long wave lengths, as long as the cut is not orthogonal to [111] ($\theta \neq \pi/2$). All plots share the same colour bar.}
    \label{fig:rotate-axes}
\end{figure*}

\section{Electric and magnetic field correlators}

The zero temperature correlation functions of the electric field may be written in terms of the normalised polarisation vectors $\epsilon_r(\vec{q})$ as
\begin{align}
    C_E^{ab}(\vec{q}) &= {\langle E^a (\vec{q}) E^b (-\vec{q}) \rangle}_{T=0} \\
    &= \frac12 \sum_{r=1}^2 \omega_r(\vec{q}) 
    \epsilon_r^a(\vec{q}) \bar{\epsilon}_r^b(\vec{q}) 
    \, ,
    \label{eqn:T=0-electric-field}
\end{align}
while the magnetic field correlation functions are given by
\begin{align}
    C_B^{\mu\nu}(\vec{q}) &= {\langle B^\mu (\vec{q}) B^\nu (-\vec{q}) \rangle}_{T=0} \\
    &= \frac12 \sum_{r=1}^2 \frac{1}{\omega_r(\vec{q})}
    M[Q(\vec{q})]^{\mu}{}_{a} \epsilon_r^a(\vec{q}) \bar{\epsilon}_r^b(\vec{q}) M[Q(\vec{q})]^\dagger_{b}{}^{\nu}
    \, ,
\end{align}
where the matrix $M[Q]$ corresponds to the momentum space representation of Eq.~\eqref{eqn:M-matrix}.
The zero temperature electric and magnetic field correlation functions are shown in Figs.~\ref{fig:E-correlation} and \ref{fig:B-correlation}, respectively.
It is evident that quantum fluctuations at strictly zero temperature wash out the pinch point structure in $C_E^{ab}(\vec{q})$.
However, if temperature is raised from zero to some nonzero value, Eq.~\eqref{eqn:T=0-electric-field} is modified to
\begin{equation}
    C_E^{ab}(\vec{q}) = \frac12 \sum_{r=1}^2 \omega_r(\vec{q}) 
    \epsilon_r^a(\vec{q}) \bar{\epsilon}_r^b(\vec{q}) \coth[\tfrac12 \beta \omega_r(\vec{q})]
    \, .
\end{equation}
For sufficiently long wavelengths, $\beta \omega_r(\vec{q}) \ll 1$ for the gapless photon branch.
At such long wavelengths, it is appropriate to keep the leading term in the small argument expansion of 
$\coth(x)=1/x + x/3+\ldots$ This leads to a cancellation of the $\omega_r(\vec{q})$ factor that was responsible for washing out the pinch points:
\begin{equation}
    C_E^{ab}(\vec{q}) = T   
    \epsilon_2^a(\vec{q}) \bar{\epsilon}_2^b(\vec{q}) + \ldots
    \, ,
\end{equation}
where we have kept only the singular contribution coming from the gapless photon branch.
Therefore, the magnitude of the pinch points is proportional to temperature.
Precisely the same mechanism is at play in, for example, quantum spin ice~\cite{BentonSikoraShannon}.
The evolution of the pinch points with temperature is shown in Fig.~\ref{fig:temperature-variation},
demonstrating that they gradually fade away as temperature is lowered to zero.

The magnetic field correlation functions, $C_B^{\mu\nu}$, do not exhibit pinch points in the vicinity of $q=0$, neither at zero temperature, nor at nonzero temperatures.
Here we show explicitly that this is the case.
At nonzero temperature, the correlation functions evaluate to
\begin{equation}
    C_B^{\mu\nu}(\vec{q})
    = \frac12 \sum_{r=1}^2 \frac{\coth[\tfrac12 \beta \omega_r(\vec{q})]}{\omega_r(\vec{q})}
    M^{\mu}{}_{a} \epsilon_r^a(\vec{q}) \bar{\epsilon}_r^b(\vec{q}) M^\dagger_{b}{}^{\nu}
    \, .
    \label{eqn:magnetic-field-correlator}
\end{equation}
First, for any correlator involving $B_4$, the gapless polarisation vector $\epsilon_2(\vec{q})$ is orthogonal to $\xi_1(\vec{q})$, which is proportional to the row $\bar{M}[Q(\vec{q})]^{4}{}_{a}$, and therefore gives a vanishing contribution to~\eqref{eqn:magnetic-field-correlator}.
The remaining correlation functions, $C_B^{ab}(\vec{q})$, are plotted in Fig.~\ref{fig:B-correlation}, both at $T=0$ and $T=ca^{-1}$. In both cases the correlation functions exhibit no singular behaviour near the origin.

%%%%%%%%%%%%%%%%%%%%%%%%%%%%%%%%%%%%%%%%%%%%%%%%%%%%%%%%%%%%%%%%%%%%%
%                         TEMPORAL GAUGE                            %
%%%%%%%%%%%%%%%%%%%%%%%%%%%%%%%%%%%%%%%%%%%%%%%%%%%%%%%%%%%%%%%%%%%%%

\section{$\Phi=0$ gauge and lattice Hamiltonian}

Consider canonically conjugate variables $A_{\mathsf{i}}^{a}$ and $E_{\mathsf{i}}^{a}$, $[A_\mathsf{i}^{a}, E_\mathsf{i}^{a}]=-i$, where sans serif indices $\mathsf{i}, \mathsf{j}, \mathsf{k}, \ldots$ label the sites of a cubic lattice.
In contrast to the Coulomb gauge, the electric fields and vector potentials satisfy canonical commutation relations, but the physical Hilbert space is constrained by Gauss' law.
In the $\Phi=0$ gauge (also sometimes called the temporal~\cite{hatfield1998quantum}, Hamiltonian or Weyl gauge), there exists some residual gauge freedom; time-independent gauge transformations are still permitted~\cite{FradkinQFTIntegrated}.
This representation allows us to touch base with microscopic Hamiltonians where Gauss' law often emerges due to an energetic constraint.
In terms of these variables, the Gauss law constraint is given by
\begin{equation}
    \Delta_{a ; \mathsf{i j}}^\dagger E_\mathsf{j}^{a} = n_\mathsf{i} = 0
    \, ,
\end{equation}
where the $\Delta_{a}^\dagger$ are finite difference operators corresponding to the discretisation of the differential operators $D_a^\dagger$ introduced in the main text.
Explicitly, the finite difference operators take the form
\begin{align}
    \Delta_{1;\msf{j k}}^\dagger E^{1}_{\mathsf{k}} &= \sum_{i} 
    \left( E^{1}_{\vec{r}_\mathsf{j} + \vec{e}_i} -  E^{1}_{\vec{r}_\mathsf{j}} \right) \\
    \Delta_{2;\msf{jk}}^\dagger E^{2}_{\msf{k}} &= \sum_{i} 
    \left( E^{2}_{\vec{r}_{\msf{j}} + 2\vec{e}_i} - 2E^{2}_{\vec{r}_{\msf{j}} + \vec{e}_i} +  E^{2}_{\vec{r}_{\msf{j}}} \right) \\
    \Delta_{3;\msf{jk}}^\dagger E^{3}_{\msf{k}} &= \sum_{i<j} 
    \left( E^{3}_{\vec{r}_{\msf{j}} + \vec{e}_i + \vec{e}_j} - E^{3}_{\vec{r}_\msf{j} + \vec{e}_i} - E^{3}_{\vec{r}_\msf{j} + \vec{e}_j} +  E^{3}_{\vec{r}_{\msf{j}}} \right) 
\end{align}
where $\vec{e}_i$ are Cartesian basis vectors, $i,j,\ldots \in \{x, y, z\}$, and the lattice spacing has been set equal to unity, $a = 1$.
The canonical commutation relations between $A_\msf{i}^a$ and $E_\msf{i}^a$ imply that $e^{iA_\msf{i}^a}$
acts as a raising operator for the corresponding electric field operator, and hence
gives rise to the charge configurations in Fig.~2 in the main text.
The local operators $n_\mathsf{i}$ mutually commute, $[n_\mathsf{i}, n_\mathsf{j}]=0$, and are conserved quantities $[H, n_\mathsf{i}]=0$.
Since each $n_\msf{i}$ trivially commutes with the electric fields, the latter equality can be shown explicitly by evaluating the commutator
\begin{equation}
    [n_\mathsf{i}, g_{\mu\nu}B^\mu_\msf{j} B^\nu_\msf{j}] = 2ig_{\nu\mu} M_{\msf{jk}}^{\mu a} \Delta_{a; \msf{ki}} B^\mu_\msf{j} = 0
    \, .
\end{equation}
where, as in Eq.~\eqref{eqn:M-matrix}, $B^\mu_\msf{j} = M^{\mu a}_{\msf{jk}} A^a_\msf{k}$.
That the product $\sum_{a, \msf{k}} M_{\msf{jk}}^{\mu a} \Delta_{a; \msf{ki}}$ vanishes identically can be seen by writing the expression in matrix form
\begin{equation}
    \begin{pmatrix}
        0 & -\Delta_3 & \Delta_2 \\
        \Delta_3 & 0 & -\Delta_1 \\
        -\Delta_2 & \Delta_1 & 0 \\
        \Delta_1 & -1 & -2
    \end{pmatrix}
    \begin{pmatrix}
        \Delta_1 \\
        \Delta_2 \\
        \Delta_3 \\
    \end{pmatrix}
    =
    \begin{pmatrix}
        \Delta_2 \Delta_3 - \Delta_3 \Delta_2 \\
        \Delta_3 \Delta_1 - \Delta_1 \Delta_3 \\
        \Delta_1 \Delta_2 - \Delta_2 \Delta_1 \\
        \Delta_1^2 - \Delta_2 - 2\Delta_3
    \end{pmatrix}
\end{equation}
which vanishes by virtue of the commutativity of the finite difference operators and the nonlinear relationship $\Delta_1^2 = \Delta_2 + 2 \Delta_3$.
Infinitesimal local gauge transformations are generated by $n_\mathsf{i}$:
\begin{equation}
    e^{-i{\phi_\mathsf{i} n_\mathsf{i}}} A_\mathsf{j}^{a} e^{i{\phi_\mathsf{i} n_\mathsf{i}}} = A_\mathsf{j}^{a} + \Delta_{a; \mathsf{jk}} \phi_{\mathsf{k}}
    \, ,
\end{equation}
where $\Delta_{a; \mathsf{jk}} = \Delta^\dagger_{a; \mathsf{kj}}$ (note that since the entries of $\Delta_{a; \mathsf{jk}}$ are pure real, transposition is equivalent to Hermitian conjugation).
The physical Hilbert space is spanned by states $| \Psi \rangle$ that satisfy the Gauss law constraint $n_\mathsf{i} | \Psi \rangle = 0$, i.e., gauge-invariant states.
In microscopic lattice models, Gauss' law usually emerges from a soft energetic constraint, and working with gauge-invariant states corresponds to working in the ground state manifold of the constraint.
Indeed, this perspective was utilised in the classical spin model realisation of parabolic pinch points in the main text.
In the vector potential representation (i.e., working with eigenstates of the $A_\msf{i}^a$ operators),
the Gauss law constraint is simply
\begin{equation}
    i \Delta^\dagger_{a; \msf{jk}} \frac{\partial \Psi(\{A\})}{\partial {A_{\msf{k}}^a}} = 0
    \, ,
    \label{eqn:gauge-invariance-condition}
\end{equation}
where $\Psi(\{A\})$ is the wave function in the vector potential representation.
Motivated by conventional \U1 EM~\cite{FradkinQFTIntegrated}, we then introduce the generalised Helmholtz decomposition of the vector potential, $A_{\msf{k}}^a = (A_\perp)_{\msf{k}}^a + (A_\parallel)_{\msf{k}}^a$. The transverse component $A_\perp$ satisfies
$\Delta_{a;\msf{jk}}(A_\perp)^a_{\msf{k}} = 0$ [$A_\perp(\vec{q})$ is perpendicular to $Q(\vec{q})$ in momentum space], while the longitudinal component $A_\parallel$ [parallel to $Q(\vec{q})$ in momentum space] can be written $(A_\parallel)_\msf{k}^a = \Delta_{a;\msf{kj}}\Phi_\msf{j}$.
If we vary $\Phi_\msf{j}$, then the wave function is modified according to
\begin{equation}
    \Psi(\{A_\perp, \Phi + \delta\Phi\}) = \Psi(\{A_\perp, \Phi\}) + \delta \Phi_\msf{j} \Delta_{a;\msf{jk}}^\dagger \frac{\partial \Psi}{\partial A^a_{\msf{k}}}
    \, .
\end{equation}
Comparison with~\eqref{eqn:gauge-invariance-condition} reveals that
\begin{equation}
    \frac{\partial \Psi(\{ A \})}{\partial \Phi_\msf{j}} = 0
\end{equation}
is required for gauge invariance of the state specified by $\Psi(\{A\})$.
Therefore the space of gauge-invariant wave functions can be written as $\Psi(\{A_\perp\})$, i.e., with no parallel component of the vector potential.
Similarly, the magnetic field contribution to the Hamiltonian projects out any contribution from $A_\parallel$ (as noted previously, $Q(\vec{q})$ coincides with the null space of the matrix $M[Q(\vec{q})]$):
\begin{multline}
    \frac12 g_{\mu\nu} B^\mu(-\vec{q}) B^\nu(\vec{q}) = \\
    \frac12  M[Q(\vec{q})]_{a}^\dagger{}^{\mu} g_{\mu\nu} M[Q(\vec{q})]^{\nu}{}_{b} A_\perp^a(-\vec{q}) A_\perp^b(\vec{q})
    \, .
\end{multline}
Therefore, in momentum space, the Hamiltonian acting on gauge invariant wave functions becomes
\begin{equation}
    \frac12 \sum_\vec{q} \left[ - \frac{\partial}{\partial A_\perp^a(-\vec{q})}\frac{\partial}{\partial A_\perp^a(\vec{q})} + M_{a}^\dagger{}^{\mu} g_{\mu\nu} M^{\nu}{}_{b} A_\perp^a(-\vec{q}) A_\perp^b(\vec{q})  \right]
    \, .
    \label{eqn:gauge-invariant-H}
\end{equation}
This quadratic Hamiltonian can be diagonalised by introducing the ladder operators
\begin{align}
    a_\vec{q}^r &= \sum_{a=1}^3 \frac{\epsilon_{r}^a(\vec{q})}{\sqrt{2\omega_r(\vec{q})}} \left[ \frac{\partial}{\partial A_\perp^a(-\vec{q})} + \omega_r(\vec{q}) A_\perp^a(\vec{q}) \right] \\
    {a_\vec{q}^r}^\dagger &= \sum_{a=1}^3 \frac{\epsilon_{r}^a(-\vec{q})}{\sqrt{2\omega_r(\vec{q})}} \left[ -\frac{\partial}{\partial A_\perp^a(\vec{q})} + \omega_r(\vec{q}) A_\perp^a(-\vec{q}) \right]
    \, ,
\end{align}
where the vectors $\epsilon_r$ form an orthonormal set with $Q$: $(\epsilon_r, \epsilon_s)=\delta_{rs}$ and $(Q, \epsilon_r)=0$.
Normalisation of the polarisation vectors $\epsilon_r$ ensures that the ladder operators satisfy the canonical commutation relations $[a_\vec{q}^r, {a_\vec{k}^s}^\dagger]=\delta_{\vec{q}\vec{k}}\delta^{rs}$.
Substituting these expressions into~\eqref{eqn:gauge-invariant-H}, we find that the Hamiltonian is diagonal if the vectors $\epsilon_r$ satisfy the eigenvalue equation
\begin{equation}
    M[Q]^{\dagger a \mu} g_{\mu\nu} M[Q]^{\nu}{}_{b} \epsilon_{r}^b(\vec{q}) = \omega_r^2(\vec{q}) \epsilon_r^a(\vec{q})
    \, ,
\end{equation}
identical to~\eqref{eqn:normal-modes-evals}, leading to a simple harmonic oscillator Hamiltonian that is identical to the one derived within the Coulomb gauge [i.e., Eq.~\eqref{eqn:Coulomb-gauge-Hamiltonian}]
\begin{equation}
    H = \frac12 \sum_\vec{q}  \sum_{r=1}^{2} \omega_r(\vec{q}) \left( {a_\vec{q}^r}^\dagger a_\vec{q}^r + a_\vec{q}^r {a_\vec{q}^r}^\dagger \right)
    \, .
\end{equation}

\section{Choice of coordinates}

In Fig.~1 in the main text, and in Figs.~\ref{fig:E-correlation} and~\ref{fig:B-correlation}, we plotted the correlation functions in the plane containing the vector $[111]$. Here, we relax this constraint and consider how the pinch points appear when considering a general cut through momentum space.

If we restrict our attention to planes that intersect with the origin, there are two remaining degrees of freedom that specify the direction of the plane's normal. 
The correlation functions are rotationally invariant about $[111]$, and therefore rotating the normal about $[111]$ will leave the pinch points unchanged.
This leaves just one degree of freedom: the angle between the normal and $[111]$.
The evolution of the pinch points as a function of this angle is plotted in Fig.~\ref{fig:rotate-axes}.
We observe that the parabolic pinch points are generically present at sufficiently long wavelengths, as long as the normal to the plane does not coincide with $[111]$ (if this is the case, rotational invariance ensures that the contours will be circular, and for the specific case of $C^{12}_E(\vec{q})$ plotted in Fig.~\ref{fig:rotate-axes}, the correlation function vanishes).

%%%%%%%%%%%%%%%%%%%%%%%%%%%%%%%%%%%%%%%%%%%%%%%%%%%%%%%%%%%%%%%%%%%%%
%                           MONTE CARLO                             %
%%%%%%%%%%%%%%%%%%%%%%%%%%%%%%%%%%%%%%%%%%%%%%%%%%%%%%%%%%%%%%%%%%%%%

\section{Details of Monte Carlo simulations}

\begin{figure}[t!]
    \centering
    \subfloat[$\hat{\eta}^{(1)}_{\vec{r}}$]{\includegraphics[width=0.25\linewidth]{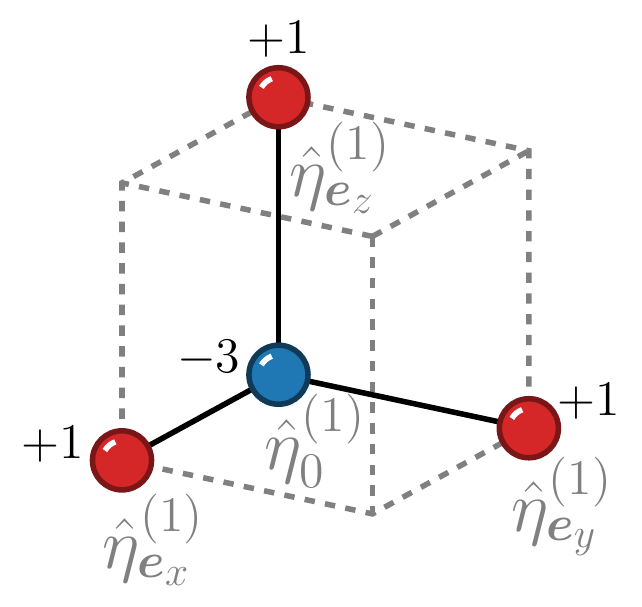}}%
    \hspace{0.015\linewidth}%
    \subfloat[$\hat{\eta}^{(2)}_{\vec{r}}$]{\includegraphics[width=0.42\linewidth]{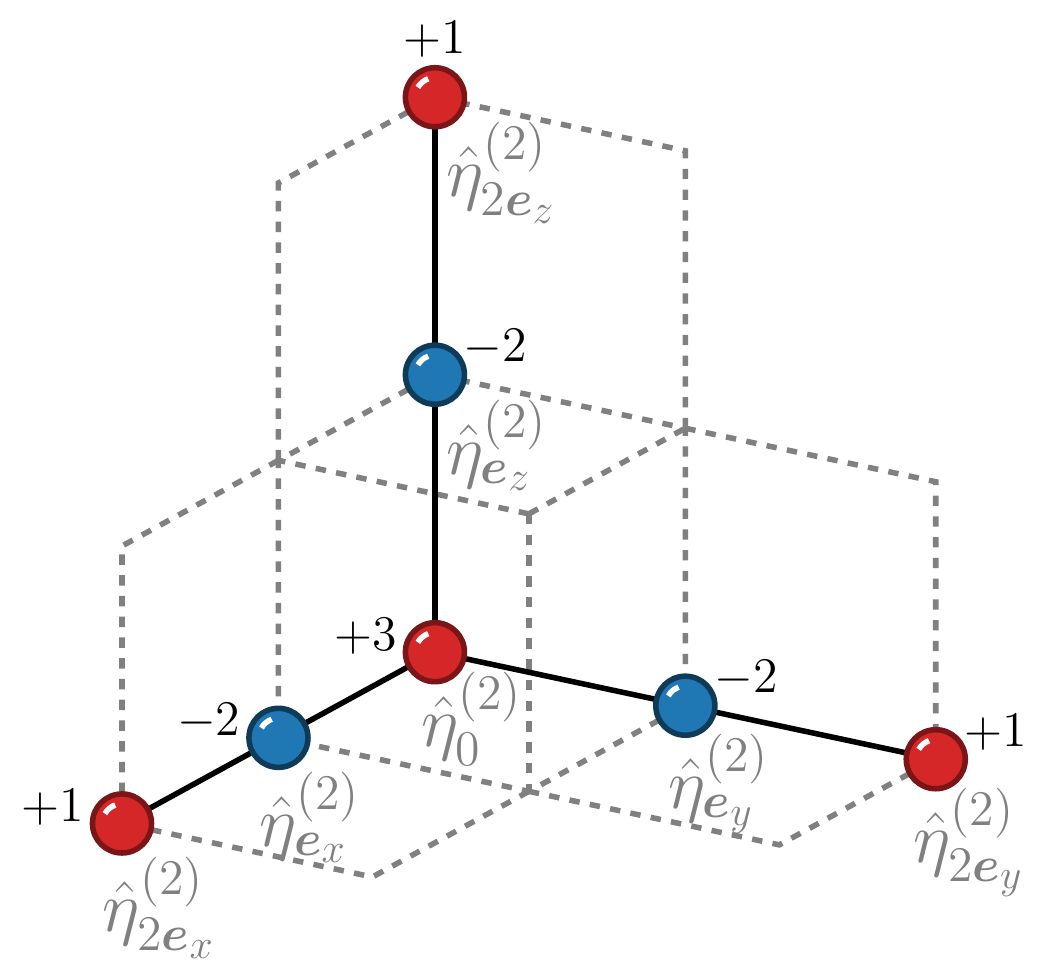}}%
    \hspace{0.015\linewidth}%
    \subfloat[$\hat{\eta}^{(2)}_{\vec{r}}$]{\includegraphics[width=0.3\linewidth]{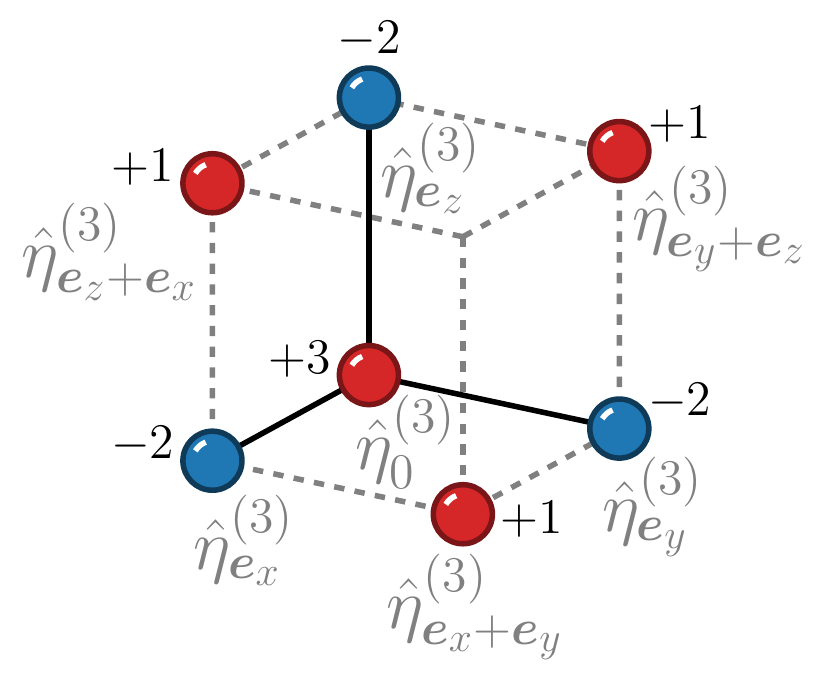}}%
    \caption{Definition of the coefficients $\hat{\eta}^{(n)}_{\vec{r}}$ that enter the classical Hamiltonian~\eqref{eqn:H-O3-verbose}.}
    \label{fig:eta-definition}
\end{figure}

In Fig.~3 in the main text, we compared the analytical large-$N$ prediction [Eq.~(20) in the main text]
with Monte Carlo simulations of the classical Hamiltonian
\begin{equation}
    H = \sum_{\vec{r}} 
    \left(  \sum_{n, \vec{r}'} \eta_{\vec{r}\vec{r}'}^{(n)} \vec{S}_\vec{r}^{(n)}  \right)^2
    \, ,
    \label{eqn:H-O3-verbose}
\end{equation}
where $\eta_{\vec{r}\vec{r}'}^{(n)} = J_n \hat{\eta}_{\vec{r}-\vec{r}'}^{(n)}$, and the coefficients $\hat{\eta}_\vec{r}$ are depicted explicitly in
Fig.~\ref{fig:eta-definition}.

Here we provide some additional details relating to these simulations.
The system is initialised in an infinite temperature state, in which each \Oh3 rotor
is uniformly distributed on the unit sphere.
The system is then evolved in time using single spin rotations and the Metropolis acceptance probability.
Specifically, a single Monte Carlo (MC) `step' consists of choosing a spin at random (both site $\vec{r}$ and sublattice $n$), and proposing
a rotation $\vec{S}_\vec{r}^{(n)}(\theta, \phi) \to \vec{S}_\vec{r}^{(n)}(\theta', \phi')$ [with no constraints on $(\theta', \phi')$
relative to $(\theta, \phi)$], where $\theta$ and $\phi$ are the polar and azimuthal angles, respectively, in a local spherical coordinate system.
The corresponding energy change $\delta E$ is computed according to the Hamiltonian~\eqref{eqn:H-O3-verbose}.
The proposed move is always accepted if the energy of the system is lowered, and it is accepted with probability $e^{-\delta E / T}$ if $\delta E > 0$.
Repeating this procedure $3 L_x L_y L_z$ times constitutes a Monte Carlo `sweep' of the system (one unit of MC time).

The MC dynamics starts at a temperature $T_\text{i}$, and is gradually lowered to $T_\text{target}$ in $N_T$ logarithmically spaced intervals.
This cooling procedure between temperatures $T_\text{i}$ and $T_\text{target}$ totals $N_\text{eq}$ MC sweeps.
The parameters $N_\text{eq}$ and $N_T$ are chosen such that the system does not fall out of equilibrium as it is cooled.
Having equilibrated the system at the target temperature $T_\text{target}$, we then measure the correlation function $\langle \vec{S}^{{(m)}}(-\vec{q}) \cdot \vec{S}^{{(n)}}(\vec{q})  \rangle$ by sampling the state of the system every $N_\text{m}$ MC sweeps. The value of $N_\text{m}$ is determined by looking at the decay of the autocorrelation function for $|\vec{S}(\vec{q})^{(n)}|^2(t)$.
The correlation function $\vec{S}^{{(m)}}(-\vec{q}, t) \cdot \vec{S}^{{(n)}}(\vec{q}, t)$ is evaluated using a fast Fourier transform from the FFTW library~\cite{fftw3}.

For Fig.~3 in the main text, we used the following parameters for the system: $J_1 = 2$, $J_2 = 0.85$, $J_3 = 1$, while for the Monte Carlo: $T_\text{i} = 100$, $T_\text{target} = 0.8$, $N_T = 100$, $N_\text{eq} = 10^5$, $N_\text{m} = 0.5 \times 10^4$.
The system size was $L_x = L_y = L_y = 32$, or $\num{98304}$ spins.
Each history was run for $10^6$ sweeps -- excluding equilibration -- totalling 200 measurements per history. The data are averaged over 16 independent histories, and over the three equivalent coordinate directions. Finally, we pad the FFT with zeros and perform a Gaussian smoothing of the resulting `oversampled' image.

\end{document}